\documentclass[preprint,showpacs,floatfix]{revtex4-1}
\usepackage{graphicx,psfrag,amsmath,amssymb,amsfonts,latexsym,color,epsf,graphpap}

\begin{document}

\title{Nonlocal effective actions in semiclassical gravity: thermal effects in stationary geometries}
\author{M. El\'ias,  F. D. Mazzitelli and L.G. Trombetta}
\affiliation{Centro At\'omico Bariloche and Instituto Balseiro, 
Comisi\'on Nacional de Energ\'\i a At\'omica, 8400 Bariloche, Argentina.}

\date{today}

\begin{abstract} 
We compute the gravitational effective action by integrating out quantum matter fields in a weak gravitational field,
using the Schwinger-Keldysh ({\it in-in}) formalism. We pay particular attention to the  role of the initial quantum 
state in the structure of the nonlocal terms in the effective action, with an eye to nonlinear completions  of the theory
that may be relevant in astrophysics and cosmology.
In this first paper we consider a quantum scalar field in thermal equilibrium, in a stationary gravitational field. We
obtain a covariant expression for the nonlocal effective action, which can be expressed in terms of the curvature tensor, the four-velocity of the thermal bath and 
the local Tolman temperature.  We discuss the connection between the results for ultrastatic and static metrics through conformal transformations,
and the main features of the thermal corrections to the semiclassical Einstein equations.
\end{abstract}
\maketitle

\section{Introduction}\label{sec:intro}
The observation of the present accelerated expansion of the universe  catalyzed a large number of theoretical speculations,  looking for natural explanations for the acceleration. The aim of many theoretical constructions is to modify  general relativity in the infrared, in order to produce an effective
cosmological constant at the late stages of the universal evolution. One kind of such modifications are nonlocal cosmological models, which have been the subject of numerous works \cite{nonlocal1,nonlocal2,nonlocal3,nonlocal4,nonlocal5,nonlocal6}. Nonlocality naturally arises when considering the backreaction of quantum fields on the evolution of the universe.  Exact calculations 
for generic spacetimes are very difficult, and therefore there have been several alternative methods to suggest viable modifications. On the one hand, 
for weak gravitational fields, the  effective action resulting from the integration of massless or massive quantum fields is very well known. Light or massless fields could introduce significant infrared modifications. Nonlinear completions of these effective actions have been considered by several authors (see \cite{nonlocal1,nonlocal2, nonlocal3} and references therein). The nonlocal structure of the effective action is of the form ${\mathcal R} F(\Box) {\mathcal R}$ where ${\mathcal R}$ denotes the components of  the Riemann tensor and  $F$ is a non-analytic  function of the D'Alembertian.   Related phenomenological proposals consider effective actions of the form $R F(\frac{1}{\Box}R) $, where $R$ is the Ricci scalar \cite{DeserWoodard}.

The usual {\it in-out} effective actions obtained after the integration of quantum matter fields produce nonlocal and noncausal equations of motion for the metric \cite{CTP}. This problem is generally avoided in the nonlocal cosmological models by using a procedure suggested many years ago by Barvinsky and Vilkovisky \cite{BarvinskyVilkovisky}, which consists in computing the effective action in Euclidean spacetime,
obtaining the field equations, and then replacing there the Euclidean propagators by retarded propagators. It was proved there that this replacement produces the correct field equations when the quantum state of the field is the ground state, and when the metric is, initially, asymptotically flat. In more general situations the use of the so called Schwinger-Keldysh or {\it in-in}/CTP formalism \cite{CTP,CTP2} becomes unavoidable, and it is in general advocated to justify the prescription of Barvinsky and Vilkovisky. Note that, in the phenomenological models, the use of the retarded propagator is just an additional prescription which is crucial
to force the reality and causality of the equations of motion. Another strong argument in favor of using the {\it in-in} formalism is that it provides more information than its Euclidean counterpart, i.e. it includes the stochastic effects. Indeed, the resulting field equations contain a stochastic noise source, whose statistical properties are given by the imaginary part of effective action, accounting for fluctuation and dissipation effects \cite{CTP2, CamposHu,Living}.

So far in the literature most of the works regarding nonlocal modifications of general relativity induced by quantum fields consider the field to be in the  vacuum state. However, under general conditions the field is expected to be in a different quantum state, the most reasonable of which being that of finite temperature. For this reason, in this paper we will discuss the dependence of the effective action of a massless field with its quantum state, focusing on stationary spacetimes in the limit of weak gravitational fields.
 Moreover, for massless fields, the most interesting astrophysical and cosmological scenarios lie in a regime where $T \gg L^{-1}$, where $T$ is the temperature and $L$ a characteristic 
curvature scale. Therefore, we will pay particular attention to the high temperature expansion of the effective action and the resulting field equations. As we will see, the initial quantum state strongly modifies the nonlocal structure. In addition to its intrinsic interest, we expect the results to be useful for new proposals of {\it in-in} nonlinear completions of the effective action that produce real and causal equations of motion.

The dependence of the nonlocal effective action with the initial quantum state of the field has been discussed in previous papers. Indeed,
Gusev and Zelnikov \cite{GusevZelnikov} computed the Euclidean effective action for a massless quantum field in a thermal state for an ultrastatic gravitational  background (ie. in adequate coordinates the metric is $g_{00}= -1, g_{0i}=0\ \text{and}\ g_{ij}=g_{ij}(\bar{x})$), using the covariant perturbation theory \cite{BarvinskyVilkovisky}.  The {\it in-in} effective action  has been computed by Campos and Hu \cite{CamposHu} (see also \cite{Paz}), in the limit of weak gravitational fields.  They obtained the effective action and the corresponding equations of motion for the metric. The technical complexity of the calculation produces cumbersome results, and therefore we found it relevant to compute the effective action  
for stationary metrics, using an alternative procedure that allows us to  find covariant expressions for effective action.
As expected, the covariant effective action will involve not only the Riemann tensor but also a vector $u^{\mu}$ that describes the four-velocity of the thermal bath,  as well as a local redefinition of the temperature. Equivalently, it can be written in terms of the Riemann tensor and the Killing vector $\kappa^\mu$ associated to the stationary geometry.

The paper is organized as follows. We present the calculation of the effective action for weak gravitational fields in Section II, 
where we obtain explicit expressions for the nonlocal kernels. We also
write the effective action in a covariant way, adapted to nonlinear completions. 
In Section III we describe the high temperature expansion of the effective action, 
showing that the leading nonlocal contribution of the real part is linear in the temperature. In Section IV we show that, for static metrics, the effective action can be derived
from the ultrastatic case,
using a conformal transformation. In Section V we write the field equations, and discuss qualitatively some features of the quantum corrections. Section VI contains the conclusions of our work. In the Appendices we include further details of the calculations.  

\section{Effective Action}

In this section we compute the in-in effective action for a massless, minimally coupled scalar field in a nearly flat spacetime. We follow 
Ref.\cite{CamposVerdaguer}, generalizing their approach to the case of a thermal state. 
The classical action for a scalar massless quantum field with minimal coupling is given by
\begin{eqnarray}
S_{m}[g_{\mu \nu},\phi] = - \frac{1}{2} \int d^4 x \sqrt{-g}\ g^{\mu \nu} \partial_{\mu} \phi \partial_\nu \phi.
\end{eqnarray}
We are using the signature $(-+++)$ for the metric,  and natural units $\hbar=c=1$.  For a weak gravitational field we write a perturbation 
around flat spacetime as
\begin{equation}
g_{\mu \nu} (x) = \eta_{\mu \nu} + h_{\mu \nu} (x)\, ,
\end{equation}
and expand the action up to quadratic order in $h_{\mu\nu}$ 
\begin{eqnarray}
S_{m} = \frac{1}{2} \int d^4x\ \phi \Bigg( \Box + V^{(1)} + V^{(2)} + ...\Bigg)\phi\, ,
\end{eqnarray}
where the perturbative operators in derivatives are
\begin{subequations}
\begin{eqnarray}
V^{(1)} &=& - \partial_\mu \bar{h}^{\mu \nu} (x) \partial_\nu - \bar{h}^{\mu \nu}(x) \partial_\mu \partial_\nu \, , \\
V^{(2)} &=&  \partial_\mu l^{\mu \nu} (x) \partial_\nu + l^{\mu \nu}(x) \partial_\mu \partial_\nu  \, ,
\end{eqnarray}
\end{subequations}
and 
\begin{subequations}
\begin{eqnarray}
\bar{h}_{\mu \nu}(x) &=& h_{\mu \nu}(x) - \frac{1}{2} h(x) \eta_{\mu \nu}\, , \\
l_{\mu \nu}(x) &=& h_\mu^{\ \alpha}(x) h_{\alpha \nu}(x) - \frac{1}{2} h(x) h_{\mu \nu}(x) + \frac{1}{8} h^2(x) \eta^{\mu \nu} - \frac{1}{4} h_{\alpha \beta}(x) h^{\alpha \beta}(x) \eta_{\mu \nu},
\end{eqnarray}
\end{subequations}
with $h(x) = h^{\alpha}_{\,\,\alpha}(x)$. The complete thermal {\it in-in} effective action  reads
\begin{eqnarray}
\Gamma &=& \frac{i}{2} Tr \Bigg[ V^{(1)}_+  (G_{++} + G_T) + V^{(2)}_+ (G_{++} + G_T) \Bigg]
- \frac{i}{2} Tr \Bigg[ V^{(1)}_-  (G_{--} + G_T) + V^{(2)}_- (G_{--} + G_T) \Bigg]
\nonumber \\
 && - \frac{i}{4} Tr\Bigg[ V_+^{(1)} (G_{++}+G_T ) V_+^{(1)} (G_{++}+G_T) - 2 V_+^{(1)} (G_{+-}+G_T) V_-^{(1)} (G_{-+} + G_T)  
\nonumber \\
&& + V_-^{(1)} (G_{--}+G_T ) V_-^{(1)} (G_{--}+G_T) \Bigg],
\end{eqnarray}
where $G_{ab}$, ($a,b=+,-$) are the usual flat space propagators in the vacuum state,  and $G_T$ the thermal contribution. Note that $G_T$ does not depend on the CTP indices $a,b$ \cite{Paz,CamposHu}. Explicitly
\begin{subequations}
 \begin{eqnarray}
\label{Gab}
G_{\pm \pm}^\beta(x,y) &=& \int \frac{d^4q}{(2\pi)^4} e^{iq(x-y)} \Bigg[ \frac{(\mp 1)}{q^2 \mp i \epsilon}  \Bigg],
\\
G_{\pm \mp}^\beta(x,y) &=& \int \frac{d^4q}{(2\pi)^4} e^{iq(x-y)} (-2 \pi i) \theta(\mp q_0) \delta(q^2),
\end{eqnarray}
and
\begin{equation}\label{GT}
G_T =  \int \frac{d^4q}{(2\pi)^4} e^{iq(x-y)} (- 2\pi i) n(|q_0|) \delta(q^2),
\end{equation}
\end{subequations}
where $n(|q_0|) = (e^{\beta |q_0|} - 1)^{-1}$ is the bosonic thermal particle distribution and $\beta$ is the inverse temperature.

In principle, one should compute the effective action in $n$-dimensions, and absorb the infinities adding appropriate counterterms. However,
in what follows we will consider only the temperature-dependent part of the effective action $\Gamma_T$, which is finite due to the presence 
of  the Bose factor $n(|q_0|)$. Therefore we can omit the counterterms and the discussion of renormalization. We split the thermal effective action as $\Gamma_T=\Gamma^{(1)}+\Gamma^{(2)}$, with
\begin{subequations}\label{gamma1y2}
\begin{eqnarray}
\Gamma^{(1)} &=& \frac{i}{2} Tr \Bigg[ (V^{(1)}_+ + V^{(2)}_+ ) G_T \Bigg]
- \frac{i}{2} Tr \Bigg[ (V^{(1)}_-   + V^{(2)}_- ) G_T \Bigg],
\label{gamma1} \\
\Gamma^{(2)} &=&- \frac{i}{2} Tr(V_+^{(1)} G_{++}V_+^{(1)}G_T)  - \frac{i}{4} Tr(V_+^{(1)}G_TV_+^{(1)}G_T) 
\nonumber \\
&&  - \frac{i}{2} Tr(V_-^{(1)} G_{--}V_-^{(1)}G_T)  - \frac{i}{4} Tr(V_-^{(1)}G_TV_-^{(1)}G_T) 
\nonumber \\
&&- \frac{i}{4} Tr(-2 V_+^{(1)} G_{+-}V_-^{(1)}G_T)  - \frac{i}{4} Tr(-2 V_+^{(1)}G_TV_-^{(1)}G_{-+}) \nonumber \\
&&- \frac{i}{4} Tr(-2 V_+^{(1)} G_{T}V_-^{(1)}G_T), \label{gamma2}
\end{eqnarray}
\end{subequations}
where $\Gamma^{(1)}$ and $\Gamma^{(2)}$ are linear and quadratic in the propagator respectively. 
The zero temperature part can be found elsewhere, for example in Ref. \cite{CamposVerdaguer}. The main nonlocal contribution is
\begin{eqnarray}
\label{logp}
\Gamma^{(NL)}_{T=0} = - \frac{1}{23040 \pi^2} \int d^4x \int d^4y && \Bigg[ 3 R_{\mu \nu \alpha \beta}(x) R^{\mu \nu \alpha \beta}(y) - R(x) R(y) \Bigg] \times 
\nonumber \\
&& \Bigg[ \int\frac{d^4p}{(2\pi)^4} e^{ip(x-y)} \log \Bigg(\frac{p^2 - i \epsilon}{\mu_0^2} \Bigg) \Bigg]
\end{eqnarray}•

Let us now elaborate on each of the two contributions to $\Gamma_T$.

\subsection{Calculation of $\Gamma^{(1)}$}

The terms proportional to $V^{(1)}$ in Eq.\eqref{gamma1} can be written in terms of $\bar{h}_{\mu\nu}$, and read
\begin{eqnarray}\label{Gamma11}
\frac{i}{2} Tr\left[ V^{(1)} G_T \right] &=& \frac{i}{2} \int d^4x \int \frac{d^4k}{(2\pi)^4} \bar{h}_{\mu \nu}(x) k^\mu k^\nu \Big[ (-2\pi i) n(|k_0|) \delta(-k_0^2+\bar{k}^2) \Big]
\nonumber \\
&=& \frac{\pi^2}{60 \beta^4} \int d^4x \Bigg[ h_{00}(x) + h_{ij}(x) \frac{\eta^{ij}}{3} \Bigg]  \, .
\end{eqnarray}
Analogously, for the terms proportional to $V^{(2)}$,  we have
\begin{eqnarray}\label{Gamma12}
\frac{i}{2} Tr\left[ V^{(2)} G_T \right]& = &- \frac{i}{2} \int d^4x \int \frac{d^4k}{(2\pi)^4} l_{\mu \nu}(x) k^\mu k^\nu \Big[ (-2 \pi i) n(|k_0|) \delta(-k_0^2 + \bar{k}^2) \Big] \nonumber\\
&=& - \frac{\pi^2}{60 \beta^4} \int d^4x  \Bigg[- \frac{h_{00}^2(x)}{2} - \frac{h_{00}(x) h_{ij}(x) \eta^{ij}}{3} 
\nonumber\\ &-&
 \frac{h_{ij}(x) h_{lm}(x)}{6} \Big( \eta^{ij} \eta^{lm} - 2 \eta^{il} \eta^{jm} \Big)+\frac{2}{3}h_{0i}(x) h_{0j}(x)\eta^{ij} \Bigg].
\end{eqnarray}
The full action $\Gamma_1$ is obtained by adding the above two equations evaluated at $h_{\mu\nu}\to h^{(+)}_{\mu\nu}$ and then subtracting the same expression evaluated at  $h_{\mu\nu}\to h^{(-)}_{\mu\nu}$.

\subsection{Calculation of $\Gamma^{(2)}$}

The calculation of $\Gamma^{(2)}$ is much more involved. Before presenting the details, it is useful to take into account that, on general grounds \cite{CTP2, CamposHu}, the quadratic part of the CTP effective action must be of the form
\begin{eqnarray}
\Gamma^{(2)}&=&\int d^4x\int d^4 y \Bigg\{ \, h^+_{\mu\nu}(x) \Big[ H^{\mu\nu\rho\sigma}(x,y)  + i N^{\mu\nu\rho\sigma}(x,y) \Big] h^+_{\rho\sigma}(y)
\nonumber\\
&-& 
2 h^+_{\mu\nu}(x) \Big[ D^{\mu\nu\rho\sigma}(x,y) + iN^{\mu\nu\rho\sigma}(x,y) \Big] h^-_{\rho\sigma}(y)
\nonumber\\ 
&-& h^-_{\mu\nu}(x) \Big[H^{\mu\nu\rho\sigma}(x,y) - iN^{\mu\nu\rho\sigma}(x,y) \Big] h^-_{\rho\sigma}(y)\, \Bigg\}, \label{gamma-hu-parametrization}
\end{eqnarray}
for some real and symmetric kernels $H(x,y)$ and $N(x,y)$, and a real and antisymmetric  kernel $D(x,y)$. The dissipation ($D$) and noise ($N$) kernels are related by a fluctuation-dissipation relation. Moreover, for the particular situation considered in this paper, a stationary geometry, there are no dissipative effects and the kernel $D$ vanishes. Therefore, it is enough to compute
the $++$ part of the effective action, and read from it the nonvanishing kernels $H$ and $N$.
Furthermore, in terms of Fourier transform of kernels, the time-independence of the metric translates into $p_0=0$.

To evaluate the different terms of Eq.\eqref{gamma2} let us first consider the trace
\begin{eqnarray}
\label{AaAa}
-\frac{i}{2} Tr\left[V_+^{(1)} G_{++}V_+^{(1)}G_T\right] &=& -  \frac{i}{2} \int dt \int  d^3x d^3y  \int \frac{d^3 p}{(2 \pi)^3} e^{i \bar{p} (\bar{x} - \bar{y})} \nonumber\\ &\times&
 h_{\mu\nu}^{(+)}(\bar{x})   A^{\mu\nu\alpha\beta}(0,\bar{p}) h_{\alpha \beta}^{(+)}(\bar{y})  \, .
\end{eqnarray}
The kernel  $A^{\mu\nu\alpha\beta}(0,\bar{p})$ reads
\begin{eqnarray}
 A^{\mu\nu\alpha\beta}(0,\bar{p})
&=& \int \frac{d^4 q}{(2\pi)^4}  \Bigg[ (-2 \pi i) \delta \Big( (q-p)^2 \Big) n(|q-p|)\Bigg]_{p_0=0} \Bigg[\frac{(-1)}{q^2 - i\epsilon} \Bigg] _{p_0=0}
\times \nonumber \\
&& \times \Bigg( \Big[\eta^{\rho \mu \tau \nu} (q-p)_\rho q_\tau \Big] \Big[ \eta^ {\lambda \alpha \sigma \beta} (q-p)_\lambda q_\sigma \Big] \Bigg)_{p_0=0}, 
\label{11}
\end{eqnarray}
where $\eta^{\rho \mu \tau \nu} = \eta^{\rho \mu} \eta^{\tau \nu} - \frac{1}{2} \eta^{\mu \nu} \eta^{\rho \tau} $.

The strategy for evaluating this kernel is as follows: we first perform the integration in $q_0$, which is trivial due to the Dirac $\delta$-function 
and the fact that for a stationary metric $p_0=0$. We then perform the angular integral in the plane perpendicular to $\bar p$, and leave the answer in terms of an integral in $k\equiv\vert\bar k\vert=|\bar{q}-\bar{p}|$, and in the remaining angular variable.  

To exemplify, we schematically show the calculation of $A^{0000}(0,\bar{p})$: 
\begin{eqnarray}
A^{0000}(0,\bar{p}) &=& \int \frac{d^4 q}{(2\pi)^4}  \Bigg[ (-2 \pi i) \delta \Big( (q-p)^2 \Big) n(|q-p|)\Bigg]_{p_0=0} \Bigg[\frac{(-1)}{q^2 - i\epsilon} \Bigg] _{p_0=0}
\times \nonumber \\
&& \times\ \Bigg( \eta^{\rho 0 \tau 0} [(q-p)_\rho q_\tau] \eta^{\lambda 0 \sigma 0} [(q-p)_\lambda q_\sigma] \Bigg)_{p_0=0} \nonumber \\
&=& \frac{i}{(2\pi)^3} \int dq_0\ d^3q\   \frac{\delta( -q_0^2 + (\bar{q}-\bar{p})^2 )\ n(|q-p|)}{-q_0^2 + \bar{q}^2 - i\epsilon}
\times \nonumber \\
&& \times\  \Bigg[\frac{ q_0^2}{2} + \frac{1}{2}  (\bar{q} - \bar{p}) \bar{q} \Bigg] 
\Bigg[\frac{ q_0^2}{2} + \frac{1}{2}  (\bar{q} - \bar{p}) \bar{q} \Bigg].
\end{eqnarray}
As mentioned above, computing the integral in $q_0$ is trivial. Then, changing to the $\bar{k} = \bar{q} - \bar{p}$ variable, we arrive at
\begin{eqnarray}
A^{0000}(0,\bar{p}) &=& \frac{i}{(2\pi)^2} \int_0^\pi d\theta \sin \theta \int_0^\infty dk\ k  \frac{n(k)}{|\bar{p}|^2 + 2 \bar{p}.\bar{k} - i\epsilon}
\times \nonumber \\
&& \Big[ k^i k^j k^l k^m \eta_{ij} \eta_{lm} + k^l k^i k^j p^m  \eta_{ij} \eta_{lm}  + \frac{1}{4}  k^i k^j p^l p^m \eta_{il} \eta_{jm} \Big].
\end{eqnarray}
At this point, we can rely on symmetry properties to simplify the tensorial structure (see Appendix~\ref{app-angular}),  and find
\begin{eqnarray}
A^{0000}(0,\bar{p}) = \frac{i}{4 \pi^2} |\bar{p}|^4 \Bigg( \frac{I_{22} }{4} + I_{31} + I_{40} \Bigg),
\end{eqnarray}
where the result is now expressed in terms of the scalar integrals $I_{\alpha \gamma}$, defined as
\begin{eqnarray}
\label{Ialphagamma}
I_{\alpha \gamma}(\bar{p}) =  \int_0^\infty dk \int_{-1}^1 d\chi \frac{k^{\alpha+1} \chi^\gamma |\bar{p}|^{- \alpha}}{|\bar{p}|^2 + 2 k |\bar{p}| \chi - i \epsilon} n(k)\, ,
\end{eqnarray}
with $\chi = \cos \theta$.

In a similar way we can calculate every other contribution of $A^{\mu \nu \alpha \beta}(\bar{p})$ (see Appendix~\ref{app-hmunu}), thus obtaining from Eq.\eqref{AaAa} that
\begin{eqnarray}
\label{Gamma2++}
&& - \frac{i}{2} Tr \Big(V_+^{(1)} G_{++}V_+^{(1)}G_T \Big) =   \frac{1}{64 \pi^5} \int dt \int  d^3x d^3y  \int d^3p\ e^{i \bar{p} (\bar{x} - \bar{y})} 
\Bigg\{ h_{00}^{(+)}(\bar{x}) p^4 \Bigg( \frac{I_{22} }{4} + I_{31} + I_{40} \Bigg)  h_{00}^{(+)}(\bar{y})
\nonumber \\
&& + h_{00}^{(+)}(\bar{x}) |\bar{p}|^4 P^{ij} \Bigg( - \frac{I_{22}}{2} - \frac{I_{31}}{2} - \frac{I_{33}}{2} + I_{40} - I_{42} \Bigg) h_{ij}^{(+)}(\bar{y})
\nonumber \\
&& + h_{ij}^{(+)}(\bar{x}) h_{lm}^{(+)}(\bar{y}) \Bigg\{ |\bar{p}|^4 P^{ij} P^{lm} \Bigg( \frac{I_{22}}{4} - \frac{I_{31}}{2} + \frac{I_{33}}{2} + \frac{I_{40}}{8} - \frac{I_{42}}{4} + \frac{I_{44}}{8} \Bigg) 
 + |\bar{p}|^4 P^{im} P^{jl} \Bigg( \frac{I_{40}}{4} - \frac{I_{42}}{2} + \frac{I_{44}}{4} \Bigg) 
\nonumber\\
&&
 +  |\bar{p}|^2 P^{il} p^{j} p^{m} \Bigg[ \Bigg( \frac{I_{20}}{2} + I_{31} \Bigg)  - \Bigg( \frac{I_{22}}{2} + I_{33} \Bigg)  \Bigg] 
+ |\bar{p}|^2 p^i p^{j} P^{lm} \Bigg[ - \frac{1}{2} \Bigg( \frac{I_{22}}{2} + I_{33} \Bigg)  \Bigg] 
\nonumber \\
&& + |\bar{p}|^2 P^{ij} p^{l} p^{m} \Bigg[ - \frac{1}{2} \Bigg( \frac{I_{22}}{2} + I_{33} \Bigg) \Bigg]
 + p^i p^{j} p^{l} p^{m} \Bigg[ \frac{1}{2} \Bigg( \frac{I_{22}}{2} + I_{33}\Bigg)  \Bigg] \Bigg\} 
+ p^i  p^{j} h_{00}^{(+)}(\bar{x}) h_{ij}^{(+)}(\bar{y}) |\bar{p}|^2\Bigg( \frac{I_{22}}{2} + I_{33}\Bigg)   
\nonumber \\
&& + h_{0i}^{(+)}(\bar{x}) h_{0j}^{(+)}(\bar{y}) |\bar{p}|^2 p^i p^j \Bigg( I_{20} + 2 I_{31}  \Bigg) 
\Bigg\},
\end{eqnarray}
where $P^{ij} = \eta^{ij} - \frac{p^ip^j}{p^2}$, $P^{i0}=P^{0i}=0$ and $P^{00} = \eta^{00} = -1$. In this expression we identify two kind of terms: those which are proportional to a pair of projectors $P_{\mu\nu}=\eta_{\mu\nu} -\frac{p_\mu p_\nu }{p^2}$ and those which are not. Since the former can be easily written in terms of the Riemann tensor and the four-velocity $u^\mu$ (see Appendix \ref{app-covariant}), it is necessary to work out the latter, which are 
all proportional to combinations of the form,
\begin{eqnarray}
\label{Igammas}
I_{\alpha \gamma} + 2 I_{(\alpha+1)(\gamma+1)} &=& \int_0^\infty d k \int_{-1}^1 d\chi \ n(k) k^{\alpha+1} \chi^{\gamma} |\bar{p}|^{-\alpha-2}
\nonumber \\
&& \frac{1}{\beta^{\alpha+2} |\bar{p}|^{\alpha+2}} \int_0^\infty dz\ n(z)\ z^{\alpha+1} \frac{[1+(-1)^\gamma]}{(1 + \gamma)}
\nonumber \\
&=& \frac{1}{(\beta |\bar{p}|)^{\alpha+2}} \Gamma(\alpha+2) \zeta(\alpha+2) \frac{[1+(-1)^\gamma]}{(1 + \gamma)}\, .
\end{eqnarray}
Note that this particular combination of integrals can be exactly evaluated, and that the result vanishes for odd values of $\gamma$.


In order to complete the calculation of the $++$ part of the effective action,  it is necessary to evaluate the term $ - \frac{i}{4} Tr(V_+^{(1)}G_TV_+^{(1)}G_T) $ in Eq.\eqref{gamma2}. Using the explicit form of the propagators,  Eqs. \eqref{Gab} and \eqref{GT}, it is easy to see that $G_T(k) = (-2 \pi i ) n(|k_0|)Im(G_{++}(k))$ then this trace is purely imaginary which can be related with the imaginary part of Eq. \eqref{Gamma2++} replacing $n \rightarrow n^2$.

It is convenient to evaluate separately the real and imaginary parts of the $++$ contribution to  $\Gamma^{(2)}$: the real part can be read from Eq.\eqref{Gamma2++}, just taking the real parts of the integrals $I_{\alpha\beta}$. On the other hand, the imaginary part can be obtained from Eqs.\eqref{gamma2} and \eqref{Gamma2++},  replacing $I_{\alpha \beta} \rightarrow \hat{I}_{\alpha \beta}$
, giving the following result 
 \begin{eqnarray}
&-& \frac{i}{2} Tr(V_+^{(1)} Im(G_{++})V_+^{(1)}G_T)  - \frac{i}{4} Tr(V_+^{(1)}G_TV_+^{(1)}G_T) = 
\frac{1}{64 \pi^5} \int dt \int  d^3x\ d^3y  \int d^3p\ e^{i \bar{p} (\bar{x} - \bar{y})} 
 \times \nonumber\\
&&  \Bigg\{ h_{00}^{(+)}(\bar{x}) h_{00}^{(+)}(\bar{y}) |\bar{p}|^4\ Im\Bigg( \frac{\hat{I}_{22} }{4} + \hat{I}_{31} + \hat{I}_{40} \Bigg) 
+ h_{00}^{(+)}(\bar{x}) h_{ij}^{(+)}(\bar{y}) |\bar{p}|^4 P_{ij}\ Im\Bigg( - \frac{\hat{I}_{22}}{2} - \frac{\hat{I}_{31}}{2} - \frac{\hat{I}_{33}}{2} \nonumber\\
&&+ \hat{I}_{40} - \hat{I}_{42} \Bigg) 
+ h_{ij}^{(+)}(\bar{x}) h_{lm}^{(+)}(\bar{y}) \Bigg[ |\bar{p}|^4 P_{ij} P_{lm}\ Im\Bigg( \frac{\hat{I}_{22}}{4} - \frac{\hat{I}_{31}}{2} + \frac{\hat{I}_{33}}{2} + \frac{\hat{I}_{40}}{8} - \frac{\hat{I}_{42}}{4} + \frac{\hat{I}_{44}}{8} \Bigg) 
\nonumber\\
&& + |\bar{p}|^4 P_{im} P_{jl}\ Im\Bigg( \frac{\hat{I}_{40}}{4} - \frac{\hat{I}_{42}}{2} + \frac{\hat{I}_{44}}{4} \Bigg) \Bigg] 
 + h_{0i}^{(+)}(\bar{x}) h_{0j}^{(+)}(\bar{y}) \Bigg[ 2 |\bar{p}|^4 P^{ij}\ Im\Bigg(\hat{I}_{40} - \hat{I}_{42} \Bigg)  \Bigg]
\Bigg\}\, ,
\end{eqnarray}

where

\begin{equation}\label{hatI}
\hat{I}_{\alpha \beta}(|\bar{p}|) =  \int dk \int_{-1}^1 d\chi \frac{k^{\alpha+1} \chi^\gamma |\bar{p}|^{- \alpha}}{|\bar{p}|^2 + 2k|\bar{p}|\chi - i \epsilon} \Big( n(k) + n(k)^2 \Big)\, .
\end{equation}

\subsection{The complete effective action: covariant form}

We now collect the previous results. The real and imaginary parts of the $++$ effective action read
\begin{equation}
Re(\Gamma_{++})= \frac{i}{2} Tr \left[ (V^{(1)}_+ + V^{(2)}_+ ) G_T \right] - \frac{i}{2} Tr\left[V_+^{(1)} Re(G_{++})V_+^{(1)}G_T\right]
\end{equation}
and
\begin{equation}
Im(\Gamma_{++})= - \frac{1}{2} Tr\left[V_+^{(1)} Im(G_{++})V_+^{(1)}G_T\right]  - \frac{1}{4} Tr\left[V_+^{(1)}G_TV_+^{(1)}G_T\right]\, .
\end{equation}
First, let us notice that throughout the different contributions to $Re(\Gamma_{++})$, Eqs.~\eqref{Gamma11}, \eqref{Gamma12} and \eqref{Gamma2++}, we find explicit terms proportional to $\beta^{-4}$. However, there will also be implicit contributions coming from the $I_{\alpha \gamma}$ integrals of Eq.~\eqref{Gamma2++} that will surface only after an expansion in high temperatures is performed. This expansion will be discussed in more detail in the next section, but for now we shall borrow these terms in order to treat them in the same standing as the explicit ones. The crucial point is that, when combining all of the contributions that go as $\beta^{-4}$, the resulting term is in fact local. This is also the case for the terms proportional to $\beta^{-2}$, which in this case are all contained within the $I_{\alpha \gamma}$'s. For this reason we will separate both types of terms from the other, truly nonlocal, contributions.

All the contributions proportional to $\beta^{-4}$ in $\Gamma_{++}$ are
\begin{eqnarray}
\label{beta-4}
Re(\Gamma_{++})^{(4)} &=&  \frac{\pi^2}{60 \beta^4} \int d^4x \Bigg[ h_{00}^{(+)}(\bar{x}) + h_{ij}^{(+)}(\bar{x}) \frac{\eta^{ij}}{3} \Bigg]  - \frac{\pi^2}{60 \beta^4} \int d^4x  \Bigg[- \frac{(h_{00}^{(+)}(\bar{x}))^2}{2} - \frac{h_{00}^{(+)}(\bar{x}) h_{ij}^{(+)}(\bar{x}) \eta^{ij}}{3} 
\nonumber\\ &&-
 \frac{h_{ij}^{(+)}(\bar{x}) h_{lm}^{(+)}(\bar{x})}{6} \Big( \eta^{ij} \eta^{lm} - 2 \eta^{il} \eta^{jm} \Big)+\frac{2}{3}h_{0i}^{(+)}(\bar{x}) h_{0j}^{(+)}(\bar{x})\eta^{ij} \Bigg]
\nonumber \\
&& + \frac{1}{480 \pi \beta^4} \int dt \int d^3x \int d^3y \int d^3p\ e^{i\bar{p}(\bar{x}-\bar{y})} 
 \Bigg[ -  \frac{1}{12} h_{ij}^{(+)}(\bar{x}) h_{lm}^{(+)}(\bar{y}) (\eta^{ij} \eta^{lm} - 2 \eta^{il} \eta^{jm}) 
\nonumber \\
&& + \frac{1}{6} \eta^{ij} h_{00}^{(+)}(\bar{x}) h_{ij}^{(+)}(\bar{y}) + h_{0i}^{(+)}(\bar{x}) h_{0j}^{(+)}(\bar{y}) \eta^{ij} \Bigg] ,
\end{eqnarray}
which, after comparing with Eq.~\eqref{raizdeg}, it is found to be the weak field expansion of 
%
\begin{eqnarray}
Re(\Gamma_{++})^{(4)} &=& \frac{\pi^2}{90 \beta^4} \int d^4x \frac{\sqrt{-g^{(+)}}}{g_{00}^{(+)\, 2}}\, . \label{ReGamma4}
\end{eqnarray}
Under a similar analysis, the terms proportional to $\beta^{-2}$ combine to form the weak field expansion of 
\begin{eqnarray}
Re(\Gamma_{++})^{(2)} &=& \frac{1}{144 \beta^2} \int d^4x \frac{\sqrt{-g^{(+)}}}{g^{(+)}_{00}} R^{(+)}(\bar{x}). \label{ReGamma2}
\end{eqnarray}
After separating both these contributions, and combining Eqs.\eqref{Gamma11}, \eqref{Gamma12}, and \eqref{Gamma2++}, we obtain 
\begin{eqnarray}
\label{ReGammafinal}
&& Re(\Gamma_{++})= \frac{\pi^2}{90 \beta^4} \int d^4x \frac{\sqrt{-g^{(+)}}}{g_{00}^{(+)\, 2}}+
 \frac{1}{144 \beta^2} \int d^4x \frac{\sqrt{-g^{(+)}}}{g^{(+)}_{00}} R^{(+)}(\bar{x})
\nonumber \\
&& + \frac{1}{8 \pi^2} \int dt \int d^3x \int d^3y \int \frac{d^3p}{(2\pi)^3}\ e^{i\bar{p}(\bar{x}-\bar{y})} \Bigg\{ h_{00}^{(+)}(\bar{x}) h_{00}^{(+)}(\bar{y}) |\bar{p}|^4A(|\bar{p}|)\nonumber \\
&& + h_{00}^{(+)}(\bar{x}) h_{ij}^{(+)}(\bar{y}) |\bar{p}|^4 P^{ij}B(|\bar{p}|)
 + h_{ij}^{(+)}(\bar{x}) h_{lm}^{(+)}(\bar{y})  |\bar{p}|^4\Big[ P^{ij} P^{lm}C(|\bar{p}|)\nonumber\\
&& + P^{im} P^{jl}D(|\bar{p}|)\Big] + h_{0i}^{(+)}(\bar{x}) h_{0j}^{(+)}(\bar{y}) |\bar{p}|^4 P^{ij}E(|\bar{p}|)\Bigg\}\, ,
\end{eqnarray}
where
\begin{subequations}
\begin{eqnarray}
A(|\bar{p}|) &=&  \frac{I_{22} }{4} + I_{31} + I_{40} - \frac{\pi^4}{10 \beta^4 |\bar{p}|^4} + \frac{\pi^2}{18 \beta^2 |\bar{p}|^2}\, , \label{A}
\\
B(|\bar{p}|) &=& - \frac{I_{22}}{2} - \frac{I_{31}}{2} - \frac{I_{33}}{2} + I_{40} - I_{42} - \frac{\pi^4}{45 \beta^4 |\bar{p}|^4} - \frac{\pi^2}{36 \beta^2 |\bar{p}|^2}\,  ,\label{B}
\\
C(|\bar{p}|) &=&   \frac{I_{22}}{4} - \frac{I_{31}}{2} + \frac{I_{33}}{2} + \frac{I_{40}}{8} - \frac{I_{42}}{4} + \frac{I_{44}}{8} + \frac{\pi^4}{90 \beta^4|\bar{p}|^4} - \frac{\pi^2}{72 \beta^2 |\bar{p}|^2} \, ,\label{C}
\\
D(|\bar{p}|) &=& \frac{I_{40}}{4} - \frac{I_{42}}{2} + \frac{I_{44}}{4} - \frac{\pi^4}{45 \beta^4 |\bar{p}|^4} + \frac{\pi^2}{72 \beta^2 |\bar{p}|^2} \label{D}\, ,
\\
E(|\bar{p}|) &=&   2 I_{40} - 2 I_{42}  - \frac{2 \pi^4}{15 \beta^4|\bar{p}|^4}  + \frac{\pi^2}{72 \beta^2 |\bar{p}|^2} \label{kernelE}\, .
\end{eqnarray}
\end{subequations}
Note that the local terms in Eq.\eqref{ReGammafinal} are multiplied by powers of the Tolman  temperature $(\beta \sqrt{|g_{00}(\bar{x})|})^{-1} = \beta(\bar{x})^{-1}$ \cite{Tolman}. These contributions have been correspondingly subtracted to the nonlocal kernels, which now will not contain any terms that go as $\beta^{-4}$ or $\beta^{-2}$ when expanded at high temperatures.

Thanks to the presence of the projectors $P_{\mu\nu}$ we can write the effective action in terms of the Riemann tensor. However, as we are considering a quantum field in a thermal state, this is not enough to obtain a covariant expression, due to the presence of a privileged direction associated to the four-velocity $u^\mu$ of the thermal bath, as well as the space-dependent Tolman temperature $\beta(\bar{x})^{-1}$. In order to be able to obtain a fully covariant expression, it is necessary to also introduce the timelike Killing vector $\kappa^\mu$ ($\kappa^2 < 0$) related to the stationary nature of the geometry \cite{killing, killing2}. In terms of this vector, we have $u^\mu = \kappa^\mu/\sqrt{-\kappa^2}$ and $g_{00} = \kappa^2$. Indeed, using Eqs. \eqref{proyectorescovariantes} and \eqref{verdaguer} we obtain
\begin{eqnarray}
\label{resultado}
Re(\Gamma_{++}) &=& \frac{\pi^2}{90} \int d^4x \frac{\sqrt{-g}}{\beta^4 \kappa^4} 
+ \frac{1}{144} \int d^4x \frac{\sqrt{-g}}{\beta^2\kappa^2} R(x)
\nonumber \\
&& + \frac{1}{8 \pi^2} \int d^4x \int d^4y \Bigg\{ 4 R_{\mu \nu}(x) u^\mu u^\nu A_1( x,y)  R_{\alpha \beta}(y)  u^\alpha u^\beta \nonumber\\
&& +2 R_{\mu \nu}(x) A_2( x,y)  u^\mu u^\nu R(y) +  R(x) A_3( x,y) R(y) 
 \nonumber\\
 && + R_{\mu \nu \alpha \beta}(x) A_4( x,y) R^{\mu \nu \alpha \beta}(y) + 4 R_{\mu \nu}(x) A_5( x,y)R_{\beta}^\nu(y) u^\mu u^\beta  \Bigg\}
\end{eqnarray}

where the  nonlocal kernels  read
\begin{subequations}
\label{kernels_nolocales_final}
\begin{eqnarray}
A_1( x,y) &=& \int \frac{d^4p}{(2\pi)^4}\ e^{ip(x-y)} \Big(A(p)-B(p)+C(p)-3D(p)+E(p) \Big)\, ,
\\
A_2( x,y) &=&  \int \frac{d^4p}{(2\pi)^4}\ e^{ip(x-y)}  \Big( B(p)-2C(p)\Big) \, ,
\\
A_3( x,y) &=&  \int \frac{d^4p}{(2\pi)^4}\ e^{ip(x-y)} C(p)\, ,
\\
A_4( x,y) &=&  \int \frac{d^4p}{(2\pi)^4}\ e^{ip(x-y)} D(p)\, ,
\\
A_5( x,y) &=&   \int \frac{d^4p}{(2\pi)^4}\ e^{ip(x-y)} \Big(E(p)-2D(p)\Big)\,, 
\end{eqnarray}
\end{subequations}
and $p=\sqrt{|p^\mu p_\mu|}$. In order to simplify the notation, we omitted the superscript $(+)$ for the metric in Eq.~\eqref{resultado}. Notice that these last expressions are valid in any coordinate system, which is evidenced by the promotion of $\bar{x} \to x$ in Eqs. \eqref{resultado} and \eqref{kernels_nolocales_final}, and $|\bar{p}| \to p$ in Eqs. \eqref{kernels_nolocales_final}.
Upon choosing the coordinates such that $\kappa^{\mu} = (1,0,0,0)$ they reduce to the former expressions for which the time-independence is manifest. However, it is worth remarking that we expect additional terms in the effective action for a metric with a generic spacetime dependence. 

We can proceed in a similar way with the imaginary part of the effective action. The result is 
 \begin{eqnarray}
 \label{resultado-im}
Im(\Gamma_{++}) &=& 
\frac{1}{8 \pi^2} \int d^4x \int d^4y 
  \Bigg\{ 4 R_{\mu \nu}(x) u^\mu u^\nu \hat{A_1}( x,y)R_{\alpha \beta}(y)  u^\alpha u^\beta 
\nonumber \\
&& +2 R_{\mu \nu}(x) \hat{A_2}( x,y)  u^\mu u^\nu R(y) 
+  R(x) \hat{A_3}( x,y) R(y) 
\nonumber\\
&& + R_{\mu \nu \alpha \beta}(x) \hat{A_4}( x,y) R^{\mu \nu \alpha \beta}(y)  + 4 R_{\mu \nu}(x) \hat{A_5}( x,y)  R_{\beta}^\nu(y) u^\mu u^\beta  \Bigg\}
\end{eqnarray}
where the imaginary kernels $\hat{A_i}(x,y)$ are obtained from the corresponding expressions in Eq.\eqref{kernels_nolocales_final} replacing 
$A, B, C, D,$ and $E$  by
\begin{subequations}
\begin{eqnarray}
\hat{A}(p) &=&  Im\Big( \frac{\hat{I}_{22} }{4} + \hat{I}_{31} + \hat{I}_{40}\Big) \, ,
\\
\hat{B}(p) &=&  Im\Big( - \frac{\hat{I}_{22}}{2} - \frac{\hat{I}_{31}}{2} - \frac{\hat{I}_{33}}{2} + \hat{I}_{40} - \hat{I}_{42} \Big)\, ,
\\
\hat{C}(p) &=&   Im\Big( \frac{\hat{I}_{22}}{4} - \frac{\hat{I}_{31}}{2} + \frac{\hat{I}_{33}}{2} + \frac{\hat{I}_{40}}{8} - \frac{\hat{I}_{42}}{4} + \frac{\hat{I}_{44}}{8} \Big)\, ,
\\
\hat{D}(p) &=& Im\Big( \frac{\hat{I}_{40}}{4} - \frac{\hat{I}_{42}}{2} + \frac{\hat{I}_{44}}{4} \Big)\, ,
\\
\hat{E}(p) &=&   2 Im\Big(\hat{I}_{40} - \hat{I}_{42} \Big)\, ,
\end{eqnarray}
\end{subequations}

The above Eqs.~\eqref{resultado} and \eqref{resultado-im} are the main results of our work. We have found a covariant, nonlocal expression for the $in-in$ effective action for stationary metrics, considering a quantum massless, minimally coupled scalar field in thermal equilibrium. The nonlocal kernels are nontrivial functions of the temperature, 
and the effective action is written in terms of the curvatures and the timelike Killing vector. Our results generalize those of Ref.\cite{GusevZelnikov}, which are valid for Euclidean ultrastatic geometries (we will describe below an interesting connection between the effective actions for ultrastatic and static situations). Regarding the work of Ref.\cite{CamposHu}, we have been able to write the effective action in a fully covariant way, valid for stationary spacetimes.  

It is interesting to remark that the local terms proportional to $\beta^{-4}$  and $\beta^{-2}$ can be interpreted as a cosmological constant and an  Einstein-Hilbert action, respectively, modified by a local temperature. The same terms also appear when computing the effective action using the Schwinger-DeWitt expansion for a massive field \cite{killing2, donoghue}.  However, for more general metrics and massless quantum fields nonlocality could arise already at these orders. 
\section{High temperature expansion}

As the structure of the nonlocal kernels is rather complex, it is worth analyzing them in the high temperature limit. The main technical point is to expand the integrals $I_{\alpha\gamma}$ in Eq.\eqref{Ialphagamma} in inverse powers of $\beta$. After a change of variables we have
\begin{eqnarray}\label{Ialphagammaz0}
I_{\alpha \gamma} = \frac{1}{2 (2z_0)^{\alpha+1}}\int_0^\infty dz \frac{z^{\alpha+1}}{e^z - 1} I_\gamma ,
\end{eqnarray}
with
\begin{eqnarray}\label{Iexp}
I_\gamma = \int_{-1}^1 d\chi \frac{\chi^\gamma}{(z_0 + z \chi - i \epsilon)},
\end{eqnarray}
and $z_0=p\beta/2$. The expansion of $I_{\alpha\gamma}$ for small $z_0$ is rather involved and is described in Appendix~\ref{hightemperatureexpansion}. For the values of $\alpha$ and $\gamma$ relevant to the evaluation of the effective action, the structure of the expansion is
\begin{equation}
I_{\alpha\gamma}=\frac{a_{\alpha\gamma}}{z_0^4}+ \frac{c_{\alpha\gamma}}{z_0^2}+
\frac{d_{\alpha\gamma}}{z_0} + e_{\alpha\gamma}\log z_0 + \mathcal{O}(z_0),
\end{equation}
where $a_{\alpha \gamma}$, $c_{\alpha \gamma}$, $d_{\alpha \gamma}$ and $e_{\alpha \gamma}$ are constants. It is noticeable the absence of a term proportional to  $\beta^{-3}$. with $\mu$

We now replace the expansions into Eq.\eqref{ReGammafinal}. The main observation is that there are no contributions proportional to $\beta^{-4}$ and $\beta^{-2}$ into the kernels, i.e. we separated them on purpose. However the $\beta^{-3}$ is not there from the beginning. For 
these reasons  the leading nonlocal contribution is proportional to $\beta^{-1}$. 
The result for this last term is
\begin{eqnarray}
Re(\Gamma_{++})^{(1)} &\approx&  \frac{1}{1024} \int d^4x \sqrt{-g} \int d^4y \sqrt{-g} \frac{1}{\beta \sqrt{-\kappa^2}}\Bigg\{ 11 R_{\mu \nu}(x) u^\mu u^\nu 
\frac{1}{ \sqrt{-\square}}R_{\alpha \beta}(y)  u^\alpha u^\beta
\nonumber\\ 
&&- 5 R_{\mu \nu}(x)\ u^\mu u^\nu \frac{1}{ \sqrt{-\square}}R(y)
 + \frac{1}{4} R(x) \frac{1}{ \sqrt{-\square}}R(y) +\frac{1}{2} R_{\mu \nu \alpha \beta}(x) \frac{1}{ \sqrt{-\square}}R^{\mu \nu \alpha \beta}(y)\nonumber\\
&&- 4 R_{\mu \nu}(x) \frac{1}{ \sqrt{-\square}}R_{\beta}^\nu(y) u^\mu u^\beta  \Bigg\}\, , \label{ReGammaHT}
\end{eqnarray}
where the nonlocal kernel is a two-point function defined as
\begin{equation} \label{box-ala-menos-n}
\frac{1}{ (\sqrt{-\square})^n} \equiv \int \frac{d^4p}{(2\pi)^4}\frac{e^{i p( x - y) } }{p^n}\, .
\end{equation}
Note that the convolution of this nonlocal kernel with a time-independent function is equivalent to the convolution of the function with the kernel  $\delta(t-t')/\sqrt{-\nabla^2}$. 
As expected on dimensional grounds, this leading contribution is a linear combination of terms of the form $\mathcal R (\beta \sqrt{-\square})^{-1} \mathcal R$ where $\mathcal R$ denotes components of the Riemann tensor eventually contracted with the four-velocity $u^{\mu}$.

The integrals $I_{\alpha\gamma}$ also have terms proportional to $\log (\beta p)$, which could potentially be nonlocal subleading corrections to the effective action. An explicit evaluation gives that these terms precisely cancel the nonlocal zero temperature contribution (see Eq.\eqref{logp}), leaving only a local remainder that goes as $\log (\beta \mu)$, with $\mu$ an arbitrary renormalization scale.

Following the same steps we find, for the imaginary part 
\begin{eqnarray}
Im(\Gamma_{++}) &=& 
\frac{1}{8 \pi^2} \int d^4x \sqrt{-g} \int d^4y \sqrt{-g} 
  \Bigg\{ 4 R_{\mu \nu}(x) u^\mu u^\nu \hat{A_1}(x,y)R_{\alpha \beta}(y)  u^\alpha u^\beta 
\nonumber \\
&& +2 R_{\mu \nu}(x) \hat{A_2}(x,y)  u^\mu u^\nu R(y) 
+  R(x) \hat{A_3}(x,y) R(y) 
\nonumber\\
&& + R_{\mu \nu \alpha \beta}(x) \hat{A_4}(x,y) R^{\mu \nu \alpha \beta}(y)  + 4 R_{\mu \nu}(x) \hat{A_5}(x,y)  R_{\beta}^\nu(y) u^\mu u^\beta  \Bigg\} \label{ImGamma}\, ,
\end{eqnarray}
where
\begin{subequations}
\label{ImKernelsHT}
\begin{eqnarray}
\hat{A}_1(x,y) &\approx& \frac{11 \pi ^5}{60 (\beta \sqrt{-\square})^5}-\frac{7 \pi ^3}{96 (\beta \sqrt{-\square})^3} + \frac{73 \pi }{768 (\beta \sqrt{-\square})^2}-\frac{11 \pi }{512 (\beta \sqrt{-\square})} \\
\hat{A}_2(x,y) &\approx& \frac{\pi ^5}{10 (\beta \sqrt{-\square})^5} - \frac{\pi^3}{16 (\beta \sqrt{-\square})^3}+\frac{11 \pi }{128 (\beta \sqrt{-\square})^2}+\frac{5 \pi }{256 (\beta \sqrt{-\square})} \\
\hat{A}_3(x,y) &\approx& \frac{\pi ^5}{60 (\beta \sqrt{-\square})^5}+\frac{\pi ^3}{32 (\beta \sqrt{-\square})^3}-\frac{37 \pi }{768 (\beta \sqrt{-\square})^2} - \frac{\pi }{512 (\beta \sqrt{-\square})} \\
\hat{A}_4(x,y) &\approx& \frac{\pi ^5}{30 (\beta \sqrt{-\square})^5} - \frac{\pi ^3}{48 (\beta \sqrt{-\square})^3}+\frac{11 \pi }{384 (\beta \sqrt{-\square})^2}-\frac{\pi }{256 (\beta  \sqrt{-\square})} \\
\hat{A}_5(x,y) &\approx& \frac{\pi ^5}{5 (\beta \sqrt{-\square})^5}-\frac{\pi ^3}{24 (\beta \sqrt{-\square})^3}+\frac{3 \pi }{64 (\beta \sqrt{-\square})^2}+\frac{\pi }{128 (\beta  \sqrt{-\square})} \, .
\nonumber \\ 
\end{eqnarray}
\end{subequations}
Note that the nonlocal kernels  $\hat{A_i}(x,y)$ are linear combinations of $(\beta\sqrt{-\square})^{-n}$.

\section{Connection with the effective action for ultrastatic spacetimes}

The structure of the effective action $\Gamma[g_{\mu \nu}]$ for a static metric $g_{\mu \nu}$ (i.e. $g_{0i} = 0$) can in fact be recovered from the effective action $\bar{\Gamma}[\bar{g}_{\mu \nu}, \Omega]$ for an ultrastatic metric $\bar{g}_{\mu \nu}$ (also $\bar{g}_{00} = \eta_{00}$) through an appropriate conformal transformation 
\begin{equation}
 g_{\mu \nu}(\bar{x}) = \Omega^2(\bar{x}) \bar{g}_{\mu \nu}(\bar{x}), \label{conformal-transf}
\end{equation}
where $\Omega^2 = |g_{00}|$. However, since the action is not conformally invariant, the corresponding effective action $\bar{\Gamma}[\bar{g}_{\mu \nu}, \Omega]$ for the ultrastatic metric is associated to a different operator. Indeed, if we consider a static metric in Euclidean signature, we have for the Euclidean effective action
\begin{eqnarray}
 e^{- \Gamma[g_{\mu \nu}]} = \int \mathcal{D} \phi \, e^{- S[\phi, g_{\mu \nu}]} &\sim& \det \left( - \square + m^2 + \xi R \right)^{-1/2}, 
\end{eqnarray}
where $\xi$ is the coupling between scalar curvature and the field. Upon applying the transformation Eq.\eqref{conformal-transf} it turns into
\begin{eqnarray}
 e^{- \Gamma[g_{\mu \nu}]} &=& J[\Omega] e^{-\bar{\Gamma}[\bar{g}_{\mu \nu}, \Omega]} \notag \\
 &\sim& J[\Omega] \det \left[ - \bar{\square} + \Omega^2 m^2 + \xi \bar{R} - \left( \xi - \frac{1}{6} \right) \Delta \bar{R} \right]^{-1/2}, 
\end{eqnarray}
where $\Delta \bar{R}$ is related to the transformation of the scalar curvature and is defined in Eq.\eqref{conformal-transf-rules}. The Jacobian $J[\Omega]$ adds a local contribution to the effective action, that produces the trace anomaly. The effective actions for Euclidean ultrastatic metrics were obtained previously up to quadratic order in the curvature $\bar{R}$, by means of heat kernel methods for a wide class of operators of the form $-\bar{\square} - \bar{P} + \frac{\bar{R}}{6}$, with $\bar{P}(x)$ a generic potential that may include the curvature as well as interactions. In particular, the effective action at finite temperature has been computed in Ref.~\cite{GusevZelnikov}, where the nonlocal structure is encoded in temperature dependent kernels $\gamma_i(\beta \sqrt{-\nabla^2})$, with $i=1,..,4$. By choosing the potential $\bar{P}$ appropriately (see Appendix~\ref{app-conformal} for further details), the effective action $\bar{\Gamma}[\bar{g}_{\mu \nu}, \Omega]$ is easily obtained, and after carefully rewriting everything in terms of the static metric $g_{\mu \nu}$, the corresponding $\Gamma[g_{\mu \nu}]$ arises.

There is an important feature of $\Gamma[g_{\mu \nu}]$ that can be deduced from this picture. At high temperatures the kernels $\gamma_i$ can be expanded in powers of $\beta^{-1}$, leading to terms in the effective action that are schematically of the form,
\begin{equation}
 \int d^4 x \int d^4 y \sqrt{-\bar{g}} \bar{\mathcal{R}}(x) \frac{1}{(\beta \sqrt{-\bar{\square}})^{n}} \sqrt{-\bar{g}} \bar{\mathcal{R}}(y), \label{generic-EA-term-US}
\end{equation}
where we are using a covariant representation that reduces to the one of Ref.~\cite{GusevZelnikov} upon specializing for an ultrastatic metric in the appropriate coordinates. The geometric objects have well known transformation rules under Eq.\eqref{conformal-transf}, some of which are summarized in Eqs.~\eqref{conformal-transf-rules}, however the nonlocal operators deserve a special treatment. Besides Eq.~\eqref{box-ala-menos-n}, and alternative definition for these operators is \cite{diego},
\begin{equation}
 \frac{1}{(\sqrt{-\bar{\square}})^{n}} = \frac{2}{\pi} \sin\left( \frac{n \pi}{2} \right) \int_{0}^{\infty} d\bar{m} \, \bar{m}^{1-n}\, \frac{1}{(-\bar{\square} + \bar{m}^2)}.
\end{equation}
The massive propagator $G_{\bar{m}}(x,y) = (-\bar{\square}^2 + \bar{m}^2)^{-1}$ is not conformally invariant. However,  up to corrections proportional to the curvature, it can be shown that in four dimensions it transforms like
\begin{equation}
 \frac{1}{(-\bar{\square} + \bar{m^2})} = \Omega(x) \frac{1}{(-\square + \bar{m^2} \Omega^{-2})} \Omega(y) + \mathcal{O}(\mathcal{R}),
\end{equation}
and therefore, upon a change of integration variable $\bar{m} \to m = \bar{m} \, \Omega^{-1}$, we can obtain the approximated transformation rule for the nonlocal operator
\begin{equation}
 (-\bar{\square})^{-\frac{n}{2}} = \Omega(x)^{2-\frac{n}{2}} (-\square)^{-\frac{n}{2}} \Omega(y)^{2-\frac{n}{2}} + \mathcal{O}(\mathcal{R}).
\end{equation}
Putting this together with the transformation rules of the other geometrical objects, it can be readily seen that a term like Eq.~\eqref{generic-EA-term-US} contributes to the effective action $\Gamma[g_{\mu \nu}]$ schematically as
\begin{equation}
 \int d^4 x \int d^4 y \frac{\sqrt{-g} \mathcal{R}(x)}{(\beta \, \Omega(x))^{n/2}} \frac{1}{( \sqrt{-\square})^{n}} \frac{\sqrt{-g} \mathcal{R}(y)}{(\beta \, \Omega(y))^{n/2}} + \mathcal{O}(\mathcal{R}^3).
\end{equation}
Going back to the case under consideration, Eq.~\eqref{conformal-transf}, this is what gives a spatial dependence to the inverse temperature parameter $\beta$,
\begin{equation} 
 \beta \to \beta(\bar{x}) = \beta \sqrt{|g_{00}(\bar{x})|},
\end{equation}
in compliance with our result Eq.~\eqref{resultado}, as well as with other works in the literature \cite{Tolman,donoghue,Dowker,otros-beta-x}.

There are limitations to this approach though. On the one hand, only a static metric can be related to an ultrastatic through a conformal transformation, leaving stationary metrics (i.e $g_{0i} \neq 0$) out of this treatment. Indeed, the kernel $E(x,y)$ of Eq.~\eqref{kernelE} will never arise from this approach. 

On the other hand, it is necessary to know the ultrastatic effective action $\bar{\Gamma}[\bar{g}_{\mu \nu}, \Omega]$ for a more general type of operator than the original one whose static effective action $\Gamma[g_{\mu \nu}]$ we aim to find. This might prove equally or more challenging than a direct calculation of $\Gamma[g_{\mu \nu}]$, even for more general metrics. In our case, the known results from Ref.~\cite{GusevZelnikov}, which are able to accommodate a generic operator, were obtained for a Euclidean ultrastatic metric and therefore allows us only to obtain the equivalent Euclidean effective action for a static metric, which we can then verify to be equal to the real part of the CTP effective action $Re(\Gamma[g_{\mu \nu}])$ we computed in Eq.\eqref{resultado}, after the high temperature expansion of Eq.~\eqref{ReGammaHT}. This means that any imaginary contribution $Im(\Gamma[g_{\mu \nu}])$, i.e Eq.\eqref{ImGamma}, is absent, and thus we are missing the information on the stochastic  properties of the system. In order to obtain this last part from a conformal transformation, we would need to know the CTP equivalent of Ref.~\cite{GusevZelnikov}, that is, the CTP effective action for an ultrastatic metric associated to a generic operator, but this is likely more difficult than to perform the full CTP calculation we presented in the previous sections.

\section{Field Equations in the High Temperature limit}

In this section we compute the semiclassical Einstein equations by directly taking the variation of $S_g[g_{\mu \nu}] + S_m[g_{\mu \nu}] + \Gamma[g_{\mu\nu}]$ with respect to the metric, where $S_g$ and $S_m$ are the gravitational action and the action for classical matter sources respectively. The semiclassical Einstein equations read
\begin{eqnarray}
M_{pl}^2 G_{\mu\nu} = T^{cl}_{\mu\nu} + \langle T_{\mu\nu} \rangle + j_{\mu \nu}, \label{SEE}
\end{eqnarray}
where $M_{pl}^2$ is the Planck mass, $G_{\mu \nu}$ is the Einstein tensor, $T_{\mu\nu}$ is the stress-energy tensor  and $j_{\mu \nu}$ a stochastic noise source, i.e. $\langle j_{\mu \nu} \rangle = 0$. The stress-energy tensor of the classical matter sources $T^{cl}_{\mu\nu}$ is related with $S_m$ in the usual way, and analogously, for the expectation value of the stress-energy tensor of the quantum field we have
\begin{eqnarray}
 \langle T_{\mu\nu} \rangle = - \frac{2}{\sqrt{-g}} \frac{\delta Re( \Gamma_{++}[g_{\mu\nu}])}{\delta g_{+}^{\mu\nu}} \Bigg|_{g_{+}^{\mu\nu} = g_{-}^{\mu\nu}},
\end{eqnarray}
where it has already been taken into account the fact that the {\it in-in} formalism ensures the equations of motion are real, and thus it is only necessary to consider the variation of the real part of $\Gamma[g_{\mu\nu}]$. Furthermore, for time-independent metrics we have $Re(\Gamma_{+-}[g_{\mu\nu}]) = Re(\Gamma_{-+}[g_{\mu\nu}]) = 0$, as already discussed after Eq.~\eqref{gamma-hu-parametrization}. 

We compute $\langle T_{\mu\nu} \rangle$ by explicitly varying $Re( \Gamma_{++}[g_{\mu\nu}])$ from Eq.~\eqref{ReGammafinal} with respect to the metric. In this section we choose to work in coordinates for which the stationarity of the metric is explicit and it does not depend on the time coordinate $t$, and thus $\frac{1}{\sqrt{-\square}} \to \frac{1}{\sqrt{-\nabla^2}} \delta(t-t')$. First let us consider the local contributions proportional to $\beta^{-4}$ and $\beta^{-2}$, whose variations can be easily obtained from their nonlinear expressions. Indeed, from Eq.~\eqref{ReGamma4} for $\beta^{-4}$ we have
\begin{eqnarray} \label{Tmunu4}
 \langle T_{\mu\nu} (\bar{x}) \rangle^{(4)} &=& \rho(\bar{x}) u_{\mu} u_{\nu} + p(\bar{x}) \left( g_{\mu\nu} + u_{\mu} u_{\nu} \right),
\end{eqnarray}
where $\rho(\bar{x}) = \frac{\pi^2}{30 \beta(\bar{x})^4}$, $p(\bar{x}) = \frac{\rho(\bar{x})}{3}$ and $u^{\mu}$ are, respectively,  the energy density, pressure and four-velocity of an ideal radiation fluid with space-dependent temperature given by the Tolman temperature $\beta(\bar{x})^{-1}$. It is worth noting that a source of this form in the Einstein equations does not allow a static solution that also satisfies sensible boundary conditions (such as decaying at spatial infinity). This has been thoroughly studied in the literature, in particular regarding the stability of hot flat space \cite{Dowker}. We will not dwell on this subject, but just note that a full treatment requires to consider a time-dependent scenario.

Similarly, from Eq.~\eqref{ReGamma2} for $\beta^{-2}$ we obtain,
\begin{eqnarray} \label{Tmunu2}
 \langle T_{\mu\nu} (\bar{x}) \rangle^{(2)} &=& \frac{1}{72} \left[ \frac{1}{\beta(\bar{x})^2} \left( G_{\mu\nu} - R\, u_{\mu} u_{\nu} \right) - \left( \nabla_\mu \nabla_\nu - g_{\mu\nu} \square \right)\frac{1}{\beta(\bar{x})^2} \right].
\end{eqnarray}
The first parenthesis will have the effect of ``renormalizing'' Newton's constant by a finite, temperature and space-dependent amount, but in a noncovariant way, i.e. the `$00$' equation receives a different correction than the others. The second parenthesis survives due to the fact that $\nabla_\mu \beta(\bar{x}) \neq 0$.

We now consider the nonlocal part (NL), which starts at $\beta^{-1}$. We take the variation from its expression in terms of $h_{\mu\nu}$, given in Eq.~\eqref{ReGammafinal} and valid when $\vert h_{\mu\nu}\vert \ll 1$, and then rewrite the result in terms of the components of the Riemann tensor. This naturally separates the different components of the equations,
\begin{subequations}
\label{TmunuNL}
\begin{eqnarray} 
\langle T_{00}(\bar{x}) \rangle^{(NL)} &=& - \frac{1}{4 \pi^2} \int d^4y \Bigg[ \Big( 4 A(t,t';\bar{x},\bar{y}) + 2 B(t,t';\bar{x},\bar{y}) \Big) \nabla^2 R_{\mu \nu}(\bar{y}) u^\mu u^\nu 
\nonumber \\
&& + B(t,t';\bar{x},\bar{y}) \nabla^2 R(\bar{y}) \Bigg]
\\
\langle T_{ij}(\bar{x}) \rangle^{(NL)} &=& - \frac{1}{2 \pi^2} \int d^4y \Bigg[ \Bigg(\eta_{ij} \nabla^2  - \nabla_i \nabla_j \Bigg) \Bigg( (B(t,t';\bar{x},\bar{y})+ 2 C(t,t';\bar{x},\bar{y})) R_{\mu \nu}(\bar{y}) u^\mu u^\nu 
\nonumber \\
&& + C(t,t';\bar{x},\bar{y}) R(\bar{y}) \Bigg)  + 2 D(t,t';\bar{x},\bar{y}) \nabla^2 R_{ij}(\bar{y})  -  D(t,t';\bar{x},\bar{y}) \nabla_i \nabla_j R(\bar{y}) \Bigg]
\\
\langle T_{0i}(\bar{x}) \rangle^{(NL)} &=& - \frac{1}{\pi^2} \int d^4y \, E(t,t';\bar{x},\bar{y}) \nabla^2 R_{\mu i}(\bar{y}) u^\mu
\end{eqnarray}
\end{subequations}
where $\langle T_{\mu \nu} \rangle$ is explicitly dependent only on $\bar{x}$, and
\begin{subequations}
\begin{eqnarray}
A(t,t';\bar{x},\bar{y}) &\approx& \delta(t-t') \left[ \frac{\pi ^2}{64 \beta \sqrt{-\nabla^2}}+\frac{1}{120} \log \left(\beta \mu \right) \delta^{(3)}(\bar{x}-\bar{y}) \right] , 
\\
B(t,t';\bar{x},\bar{y}) &\approx& \delta(t-t') \left[- \frac{\pi ^2}{64 \beta \sqrt{-\nabla^2}} - \frac{1}{80} \log \left(\beta \mu \right) \delta^{(3)}(\bar{x}-\bar{y}) \right], 
\\
C(t,t';\bar{x},\bar{y}) &\approx& \delta(t-t') \left[ \frac{\pi ^2}{512 \beta \sqrt{-\nabla^2}} + \frac{1}{160} \log \left( \beta \mu \right) \delta^{(3)}(\bar{x}-\bar{y}) \right], 
\\
D(t,t';\bar{x},\bar{y})&\approx& \delta(t-t') \left[ \frac{\pi ^2}{256 \beta \sqrt{-\nabla^2}}+\frac{1}{480} \log \left(\beta \mu\right) \delta^{(3)}(\bar{x}-\bar{y}) \right], 
\\
E(t,t';\bar{x},\bar{y}) &\approx& - \frac{1}{480} \log \left(\beta \mu \right) \delta(t-t') \delta^{(3)}(\bar{x}-\bar{y}).
\end{eqnarray}
\end{subequations}
Notice that the logarithmic term is local due to the cancellation of all $\log(p)$ parts with the usual zero temperature contributions (see Eq. \eqref{logp}). Also, $\langle T_{0i}(\bar{x}) \rangle^{(NL)}$ is purely local. 

The expressions in Eqs.~\eqref{TmunuNL} show only contributions linear in the curvature to $\langle T_{\mu\nu} \rangle^{(NL)}$, of the form $\int (\sqrt{-\nabla^2})^{-1} \nabla^2 \mathcal{R}$. This is due to the level of approximation used when computing the effective action, quadratic in the metric perturbation $h_{\mu\nu}$, and thus linear in the field equations. If we consider any of the possible non-linear completions of the effective action, its variation will contain further terms of higher order in the curvature $\mathcal{R}$ in the nonlocal part of the field equations, of the form
\begin{eqnarray}
 \int \mathcal{G}_{\mu\nu} \frac{1}{\sqrt{-\nabla^2}} \mathcal{R} + \int \mathcal{R} \frac{\delta}{\delta g^{\mu \nu}} \left( \frac{1}{\sqrt{-\nabla^2}} \right) \mathcal{R},
\end{eqnarray}
where $\mathcal{G}_{\mu\nu}$ represents a generic variation of components of the Riemann tensor with respect to the metric.

The full $\langle T_{\mu\nu} \rangle$ is obtained by adding all the previous contributions,
\begin{eqnarray}\label{full-Tmunu}
\langle T_{\mu\nu} \rangle = \langle T_{\mu\nu} \rangle^{(4)} + \langle T_{\mu\nu} \rangle^{(2)} + \langle T_{\mu\nu} \rangle^{(NL)}.
\end{eqnarray}
From the explicit expressions for each contribution it can be checked that the full stress-tensor is conserved, i.e. $\nabla_\mu \langle T^{\mu \nu} \rangle= 0$.

Let us now discuss the stochastic contribution to Eq.~\eqref{SEE}. The noise $j_{\mu \nu}$ accounts for the fluctuations of the stress-energy tensor of the quantum field around its mean value. Its self-correlation is related to the imaginary part of the effective action,
\begin{eqnarray}
 \langle j_{\mu\nu}(x) j_{\rho\sigma}(y) \rangle = \langle \{ \Delta T_{\mu \nu}(x) ; \Delta T_{\rho \sigma}(y) \} \rangle = N_{\mu \nu \rho \sigma}(x,y),
\end{eqnarray}
where $\Delta T_{\mu \nu} = T_{\mu\nu} - \langle T_{\mu\nu} \rangle$, and the noise kernel $N_{\mu \nu \rho \sigma}(x,y)$ is defined in Eq.~\eqref{gamma-hu-parametrization}. The semiclassical approximation will be a good one as long as the fluctuations of the stress-energy tensor are small compared to its mean value, that is,
\begin{eqnarray}
 \langle \Delta T_{\mu \nu} ^2 \rangle \ll \langle T_{\mu \nu} \rangle^2.
\end{eqnarray}
Note that since $j_{\mu\nu}(x)$ is of stochastic origin, it depends both on space and time coordinates. This can be made compatible with a static metric if the dominant source of Eq.~\eqref{SEE} is the semiclassical one and the solution is stable under small time-dependent perturbations. 



We close this section by considering the ansatz of a static Newtonian metric, having the form 
\begin{eqnarray}
 ds^2 = -(1+2 \phi(\bar{x})) dt^2 + (1 - 2\phi(\bar{x}))\ d\bar{x}^2.
\end{eqnarray}
We first assume there is a mechanism by which the effect of the source $\langle T_{\mu\nu} \rangle^{(4)}$ is stabilized such that a static metric can be a viable solution. The most striking feature is that for this kind of metric the leading nonlocal contribution $\langle T_{\mu\nu} \rangle^{(NL)}$ proportional to $\beta^{-1}$ vanishes exactly, although this can be checked to be a special feature of the minimally coupled case ($\xi = 0$), and not hold for more general cases. In the absence of both $\langle T_{\mu\nu} \rangle^{(4)}$ and $\langle T_{\mu\nu} \rangle^{(NL)}$,  the noise source $j_{\mu\nu}$ becomes dominant. This would break the semiclassical approximation for Newtonian metrics at high temperatures.


\section{Conclusions}

We have  computed the CTP effective action for a massless quantum field in thermal equilibrium in a stationary background. Our main goal has been to obtain covariant expressions for the real and imaginary parts of the effective action, which may be used to propose nonlinear completions of the theory
that are much more general than previous ones, based on generalizations of the zero temperature case. The existence of a thermal bath picks up a particular frame, the one in which the bath is at rest, and therefore the effective action depends not only on the metric but also on the four-velocity of the bath and 
on the local Tolman temperature. The nonlocal terms involve kernels that can be thought schematically as non-analytic functions of the Laplacian $F(\nabla^2)$.

The high temperature expansion revealed some interesting properties. On the one hand, and similarly to what happens in the ultrastatic case 
\cite{GusevZelnikov}, the leading nonlocal correction to the real part of the effective action is proportional to $\beta^{-1}$. By dimensional analysis, this implies that the nonlocal structure should involve the kernel $1/\sqrt{-\nabla^2}$. On the other hand, the imaginary part contains nonlocal contributions
already at the $\beta^{-5}$-level. 

Another interesting aspect of our results is that the leading nonlocal contributions to the semiclassical Einstein equations vanish when evaluated
in a Newtonian metric, and therefore do not produce thermal corrections in the large distance limit. However, we have checked that this is true
only for minimally coupled scalar fields.

In a forthcoming paper we will address the calculation of the effective action for the case of time-dependent metrics.  On general grounds, we expect the nonlocal
terms found in the present paper to become more complex functions $F(\nabla^2,u^\mu D_\mu)$,  which involve both spatial and temporal derivatives, in a combination 
that does not necessarily coincide with the D'Alembertian. An interesting issue,  with potential applications to the dark energy problem,  is whether there are nonlocal
contributions at the leading $\beta^{-4}$ order or not. We have shown that this is not the case for stationary metrics,  since in this case the leading term is local and proportional
to the fourth power of the Tolman temperature.  The same local term appears when considering massive fields, and the effective action is computed using a Schwinger-DeWitt expansion \cite{killing2,donoghue}, which assumes that the mass $m$ satisfies $m^2\gg{\mathcal R}$. However, it is, in principle,  possible that nonlocal terms emerge at leading order  in the opposite limit of massless fields, for time dependent metrics. 

\newpage

\appendix

\section{Angular Integrals} \label{app-angular}
In this Appendix we include some useful identities for the angular integrals needed to compute the quadratic part of the effective action:
\begin{eqnarray}
\label{angulares}
 \int d^3k f(k,p) k^i k^j &=& p^i p^j \Bigg[ \int d^3k\ f(|k|,|\bar{p}|,\chi) \chi^2 \Bigg] \nonumber\\
 &+& |\bar{p}|^2 P^{i j} \Bigg[ \int d^3k\ f(|k|,|\bar{p}|,\chi) \frac{1}{2} (1-\chi^2) \Bigg]\\
 \int d^3k f(k,p) k^i k^j k^l &=& p^i p^j p^l \Bigg[ \int d^3k\ f(|k|,|\bar{p}|,\chi) \chi^3 \Bigg] 
\nonumber \\
	&&+ \Bigg(P^{i j} p^l + P^{i l} p^j + P^{l j} p^i \Bigg) \Bigg[ \int d^3k\ f(|k|,|\bar{p}|,\chi) \frac{1}{2} \chi (1 - \chi^2) |\bar{p}|^2 \Bigg]
 \\
 \int d^3k f(k,p) k^i k^j k^l k^m 
	&=& p^i p^j p^l p^m \Bigg[ \int d^3k\ f(|k|,|\bar{p}|,\chi) \chi^4 \Bigg] + |\bar{p}|^2 \Bigg( P^{i j} p^l p^m + P^{i l} p^j p^m + P^{i m} p^j p^l 
\nonumber \\
	&+& P^{j l} p^i p^m + P^{j m} p^i p^l + P^{l m} p^i p^j \Bigg) \Bigg[ \int d^3k\ f(|k|,|\bar{p}|,\chi) \frac{1}{2} \chi^2 (1 - \chi^2) \Bigg]
\nonumber \\
	&+&  \Bigg( P^{i j} P^{l m} + P^{i l} P^{j m} + P^{i m} P^{j l} \Bigg) \Bigg[ \int d^3k\ f(|k|,|\bar{p}|,\chi) \frac{1}{8} (1 - \chi^2)^2 \Bigg]
\end{eqnarray}•
where $\chi = \cos \theta$ and $P^{\mu \nu}=\eta^{\mu \nu} - p^\mu p^\nu/p^2$ with $p_0=0$.

\section{Detailed calculation of $h_{\mu \nu}^{(+)}(\bar{x}) A^{\mu \nu \alpha \beta}(\bar{x},\bar{y}) h_{\alpha \beta}^{(+)}(\bar{y})$}
\label{app-hmunu}

Here we show details about the calculation of $J_{++}\equiv h_{\mu \nu}^{(+)}(\bar{x}) A^{\mu \nu \alpha \beta}(\bar{x},\bar{y}) h_{\alpha \beta}^{(+)}(\bar{y})$.

We have
\begin{eqnarray}
J_{++}&=& \int \frac{d^4 q}{(2\pi)^4}  \Bigg[ (-2 \pi i) \delta \Big( (q-p)^2 \Big) n(|q-p|)\Bigg]_{p_0=0} \Bigg[\frac{(-1)}{q^2 - i\epsilon} \Bigg] _{p_0=0}
\times \nonumber \\
&& \times\ h_{\mu \nu}^{(+)}(\bar{x}) \Bigg( \eta^{\rho \mu \tau \nu} [(q-p)_\rho q_\tau] \eta^{\lambda \alpha \sigma \beta} [(q-p)_\lambda q_\sigma] \Bigg)_{p_0=0} h_{\alpha \beta}^{(+)}(\bar{y})
\nonumber \\
&=& \frac{i}{(2\pi)^3} \int dq_0\ d^3q\   \frac{\delta( -q_0^2 + (\bar{q}-\bar{p})^2 )\ n(|q-p|)}{-q_0^2 + \bar{q}^2 - i\epsilon}
\times 
\nonumber \\
&& \times\  \Bigg\{ h_{00}^{(+)}(\bar{x}) h_{00}^{(+)}(\bar{y}) \Bigg[\frac{ q_0^2}{2} + \frac{1}{2}  (\bar{q} - \bar{p}) \bar{q} \Bigg] 
\Bigg[\frac{ q_0^2}{2} + \frac{1}{2}  (\bar{q} - \bar{p}) \bar{q} \Bigg] 
\nonumber \\
&& + h_{00}^{(+)}(\bar{x}) h_{lm}^{(+)}(\bar{y})\Bigg[\frac{ q_0^2}{2} + \frac{1}{2}  (\bar{q} - \bar{p}) \bar{q} \Bigg] 
\Bigg[(q-p)^l q^m - \frac{1}{2} \eta^{lm} \Big(-q_0^2 + (\bar{q} - \bar{p}) \bar{q} \Big) \Bigg] 
\nonumber \\
&& + h_{ij}^{(+)}(\bar{x}) h_{00}^{(+)}(\bar{y}) \Bigg[(q - p)^i q^j - \frac{1}{2} \eta^{ij} \Big(-q_0^2 + (\bar{q} - \bar{p}) \bar{q} \Big) \Bigg] 
\Bigg[\frac{ q_0^2}{2} + \frac{1}{2}  (\bar{q} - \bar{p}) \bar{q} \Bigg] 
\nonumber \\
&& + h_{ij}^{(+)}(\bar{x}) h_{lm}^{(+)}(\bar{y}) \Bigg[(q - p)^i q^j - \frac{1}{2} \eta^{ij} \Big(-q_0^2 + (\bar{q} - \bar{p}) \bar{q} \Big) \Bigg] 
\nonumber \\
&&
\times \Bigg[(q - p)^l q^m - \frac{1}{2} \eta^{lm} \Big(-q_0^2 + (\bar{q} - \bar{p}) \bar{q} \Big) \Bigg] 
\nonumber \\
&& + h_{0i}^{(+)}(\bar{x}) h_{0j}^{(+)}(\bar{y}) q_0^2 \Bigg[ q^i + (q-p)^i \Bigg] \Bigg[q^j + (q-p)^j \Bigg] \Bigg\}
\end{eqnarray}

The integral in $q_0$  is trivial because of the Dirac $\delta$-function.  Then, changing variables as $\bar{k} = \bar{q} - \bar{p} \equiv k$ and performing the integral
in $\phi$ we obtain
\begin{eqnarray}
J_{++}&=& \frac{i}{(2\pi)^2} \int_0^\pi d\theta \int_0^\infty dk\ k  \frac{n(k)}{|\bar{p}|^2 + 2 \bar{p}.\bar{k} - i\epsilon} \times \Bigg\{ h_{00}^{(+)}(\bar{x}) h_{00}^{(+)}(\bar{y}) \Bigg[\bar{k}^2 + \frac{\bar{k}.\bar{p}}{2} \Bigg]^2
\nonumber \\
&& 
+ h_{00}^{(+)}(\bar{x}) h_{lm}^{(+)}(\bar{y}) \Bigg[\bar{k}^2 + \frac{\bar{k}.\bar{p}}{2} \Bigg] 
\Bigg[k^l k^m + k^l p^m - \frac{1}{2} \eta^{lm} \bar{k}.\bar{p} \Bigg] 
\nonumber \\
&& + h_{ij}^{(+)}(\bar{x}) h_{00}^{(+)}(\bar{y}) \Bigg[k^i k^j + k^i p^j - \frac{1}{2} \eta^{ij} \bar{k}.\bar{p} \Bigg] 
\Bigg[\bar{k}^2 + \frac{\bar{k}.\bar{p}}{2} \Bigg] 
\nonumber \\
&& + h_{ij}^{(+)}(\bar{x}) h_{lm}^{(+)}(\bar{y}) \Bigg[k^i k^j + k^i p^j - \frac{1}{2} \eta^{ij} \bar{k}.\bar{p} \Bigg] 
\Bigg[k^l k^m + k^l p^m - \frac{1}{2} \eta^{lm} \bar{k}.\bar{p} \Bigg] 
\nonumber \\
&& + h_{0i}^{(+)}(\bar{x}) h_{0j}^{(+)}(\bar{y}) \bar{k}^2 \Big[ 2 k^i  p^i \Big] \Big[ 2 k^j + p^j \Big] \Bigg\}
\end{eqnarray}•

We split the remaining expressions separating the terms  proportional to $k^4$, $k^3$ and $k^2$:
\begin{eqnarray}
J_{++}&=& \frac{i}{(2\pi)^2} \int_0^\pi d\theta \int_0^\infty dk\ k  \frac{n(k)}{p^2 + 2 \bar{p}.\bar{k} - i\epsilon}
\Bigg[ K^{(4)}(\bar{x},\bar{y},\bar{k}) + K^{(3)}(\bar{x},\bar{y},\bar{k}) \nonumber\\
&&+ K^{(2)}(\bar{x},\bar{y},\bar{k}) \Bigg]
\end{eqnarray}
where
\begin{eqnarray}
K^{(4)}(\bar{x},\bar{y},\bar{k}) &=& k^i k^j k^l k^m  \Bigg[ h_{00}^{(+)}(\bar{x}) h_{00}^{(+)}(\bar{y}) \eta_{ij} \eta_{lm} + h_{00}^{(+)}(\bar{x}) h_{lm}^{(+)}(\bar{y}) \eta_{ij} 
+ h_{ij}^{(+)}(\bar{x}) h_{00}^{(+)}(\bar{y}) \eta_{lm} 
\nonumber \\
&& + h_{ij}^{(+)}(\bar{x}) h_{lm}^{(+)}(\bar{y}) + 4  h_{0i}^{(+)}(\bar{x}) h_{0j}^{(+)}(\bar{y}) \eta^{lm}  \Bigg]
\nonumber \\
K^{(3)}(\bar{x},\bar{y},\bar{k}) &=& h_{00}^{(+)}(\bar{x}) h_{00}^{(+)}(\bar{y}) \eta_{ij} \eta_{lm} p^m k^l k^i k^j + h_{00}^{(+)}(\bar{x}) h_{lm}^{(+)}(\bar{y}) \Bigg[ \eta_{ij} k^i k^j k^l p^m - \frac{1}{2} \eta^{lm} \eta_{ij} k^i k^j \eta_{br} k^b p^r) 
\nonumber \\
	&& + \frac{\eta_{ij} k^i p^j}{2} k^l k^m \Bigg]  + h_{ij}^{(+)}(\bar{x}) h_{00}^{(+)}(\bar{y}) \Bigg[\eta_{lm} k^l k^m k^i p^j - \frac{1}{2} \eta^{ij} \eta_{lm} k^l k^m \eta_{br} k^b p^r + \frac{1}{2} \eta_{lm} k^l p^m k^i k^j \Bigg]
\nonumber \\
	&& + h_{ij}^{(+)}(\bar{x}) h_{lm}^{(+)}(\bar{y}) \Bigg[ k^i k^j k^l p^m - \frac{1}{2} \eta^{lm} \eta_{br} k^b p^r k^i k^j + k^l k^m k^i p^j - \frac{1}{2} \eta^{ij} k^l k^m \eta_{br} k^b p^r \Bigg] 
\nonumber \\
&& + 2 h_{0i}^{(+)}(\bar{x}) h_{0j}^{(+)}(\bar{y}) \eta_{lm} k^l k^m \Bigg[ k^i p^j + k^j p^i \Bigg]
\nonumber \\
K^{(2)}(\bar{x},\bar{y},\bar{k}) = &=& \frac{1}{4} h_{00}^{(+)}(\bar{x}) h_{00}^{(+)}(\bar{y}) \eta_{ij} \eta_{lm} k^i p^j k^l p^m
+ \frac{1}{2} h_{00}^{(+)}(\bar{x}) h_{lm}^{(+)}(\bar{y}) \eta_{ij} k^i p^j
\Bigg[k^l p^m - \frac{1}{2} \eta^{lm} \eta_{br} k^b p^r \Bigg] 
\nonumber \\
&& + \frac{1}{2} h_{ij}^{(+)}(\bar{x}) h_{00}^{(+)}(\bar{y}) \Bigg[ k^i p^j - \frac{1}{2} \eta^{ij} \eta_{br} k^b p^r \Bigg] 
\eta_{lm} k^l p^m + h_{0i}^{(+)}(\bar{x}) h_{0j}^{(+)}(\bar{y}) \eta_{lm} k^l k^m p^i p^j
\nonumber \\
&& + h_{ij}^{(+)}(\bar{x}) h_{lm}^{(+)}(\bar{y}) \Bigg[k^i p^j - \frac{1}{2} \eta^{ij} \eta_{ag} k^a p^g \Bigg] 
\Bigg[k^l p^m - \frac{1}{2} \eta^{lm} \eta_{br} k^b p^r \Bigg] 
\end{eqnarray}
Hence, using the identities of Appendix A and defining the integrals
\begin{eqnarray}
\label{Ialphagammaa}
I_{\alpha \gamma}(|\bar{p}|) =  \int_0^\infty dk \int_{-1}^1 dx \frac{k^{\alpha+1} x^\gamma |\bar{p}|^{- \alpha}}{|\bar{p}|^2 + 2|\bar{k}||\bar{p}|x - i \epsilon} n(k)\, ,
\end{eqnarray}
we obtain
\begin{eqnarray}
J_{++}&=& \frac{i}{4 \pi^2} \Bigg\{ h_{00}^{(+)}(\bar{x}) |\bar{p}|^4 \Bigg( \frac{I_{22} }{4} + I_{31} + I_{40} \Bigg)  h_{00}^{(+)}(\bar{y})
\nonumber \\
&& + h_{00}^{(+)}(\bar{x}) |\bar{p}|^4 P^{ij} \Bigg( - \frac{I_{22}}{2} - \frac{I_{31}}{2} - \frac{I_{33}}{2} + I_{40} - I_{42} \Bigg) h_{ij}^{(+)}(\bar{y})
\nonumber \\
&& + h_{ij}^{(+)}(\bar{x}) h_{lm}^{(+)}(\bar{y}) \Bigg\{ |\bar{p}|^4 P^{ij} P^{lm} \Bigg( \frac{I_{22}}{4} - \frac{I_{31}}{2} + \frac{I_{33}}{2} + \frac{I_{40}}{8} - \frac{I_{42}}{4} + \frac{I_{44}}{8} \Bigg) 
\nonumber \\
&& + |\bar{p}|^4 P^{im} P^{jl} \Bigg( \frac{I_{40}}{4} - \frac{I_{42}}{2} + \frac{I_{44}}{4} \Bigg) 
 +  |\bar{p}|^2 P^{il} p^{j} p^{m} \Bigg[ \Bigg( \frac{I_{20}}{2} + I_{31} \Bigg)  - \Bigg( \frac{I_{22}}{2} + I_{33} \Bigg)  \Bigg] 
\nonumber \\
&& + |\bar{p}|^2 p^i p^{j} P^{lm} \Bigg[ - \frac{1}{2} \Bigg( \frac{I_{22}}{2} + I_{33} \Bigg)  \Bigg] 
 + |\bar{p}|^2 P^{ij} p^{l} p^{m} \Bigg[ - \frac{1}{2} \Bigg( \frac{I_{22}}{2} + I_{33} \Bigg) \Bigg]
\nonumber \\
&& + p^i p^{j} p^{l} p^{m} \Bigg[ \frac{1}{2} \Bigg( \frac{I_{22}}{2} + I_{33}\Bigg)  \Bigg] \Bigg\} 
+ p^i  p^{j} h_{00}^{(+)}(\bar{x}) h_{ij}^{(+)}(\bar{y}) |\bar{p}|^2\Bigg( \frac{I_{22}}{2} + I_{33}\Bigg)   
\nonumber \\
&& + h_{0i}^{(+)}(\bar{x}) h_{0j}^{(+)}(\bar{y}) |\bar{p}|^2 p^i p^j \Bigg( I_{20} + 2 I_{31}  \Bigg) 
\Bigg\}
\end{eqnarray}

\section{Covariant Formulas} \label{app-covariant}

In this Appendix we relate the various expressions valid in the weak field approximation up to quadratic order in $h_{\mu\nu}$ 
with covariant objects.

Some useful expansions of local geometric quantities are 
\begin{subequations}
\begin{eqnarray}
\label{raizdeg}
\sqrt{-g}^{(2)} &\approx& 1 - \frac{1}{2} h_{00} + \frac{1}{2}h_{ij}\eta^{ij} - \frac{1}{8} h_{00}^2 - \frac{1}{4} h_{00} h_{ij} \eta^{ij} + \frac{1}{8} h_{ij} h_{lm} \Big[ \eta^{ij} \eta^{lm} - 2 \eta^{il} \eta^{jm} \Big]
\nonumber \\ 
&& + \frac{1}{2} h_{0i}h_{0j} \eta^{ij} \\
 R^{(1)} &\approx& h^{\alpha \beta},_{\alpha \beta} - h,_\alpha^{\ \alpha}  \\
 R_{\mu \nu}^{(1)} &\approx& \frac{1}{2} \eta^{\beta \delta} \Big( h_{\beta \nu,\mu \delta} - h_{\mu \nu, \beta \delta} - h_{\beta \delta,\mu \nu} + h_{\mu \delta,\beta \nu} \Big)
\\
 G_{\mu \nu}^{(1)} &\approx&   R_{\mu \nu}^{(1)} - \frac{1}{2} \eta_{\mu \nu} R^{(1)},
 \label{deltaGmunu}
\end{eqnarray}
\end{subequations}

For nonlocal kernels that involve the product of projectors $P^{\mu\nu}$ we have \cite{CamposVerdaguer},
\begin{eqnarray}
\label{verdaguer}
\int d^4x\ d^4y\ R(x)\ K(x-y)\ R(y) &=& \int d^4x\ d^4y\ h_{\mu \nu}(x) h_{\alpha \beta}(y) \times
\nonumber \\
&& \int d^4p\ e^{ip(x-y)} p^4 \tilde{K}(p) P^{\mu \nu} P^{\alpha \beta} 
\nonumber \\
\int d^4x\ d^4y\ R_{\mu \nu \alpha \beta}(x)\ K(x-y)\ R^{\mu \nu \alpha \beta}(y) &=& \int d^4x\ d^4y\ h_{\mu \nu}(x) h_{\alpha \beta}(y) \times 
\nonumber \\
&& \int d^4p\ e^{ip(x-y)} p^4 \tilde{K}(p) P^{\mu \beta} P^{\nu \alpha} \, ,
\end{eqnarray}
where $\tilde K(p)$ is the Fourier transform of $K(x-y)$.
Schematically, the projectors are replaced following the rules
\begin{subequations}\label{repP}
\begin{eqnarray}
&& p^4 h_{\mu \nu}(x) h_{\alpha \beta}(y) P^{\alpha \beta} P^{\mu \nu} \rightarrow R(x) R(y)\\
&&  p^4 h_{\mu \nu}(x) h_{\alpha \beta}(y) P^{\mu \beta} P^{ \nu\alpha} \rightarrow  R_{\mu \nu \alpha \beta}(x) R^{\mu \nu \alpha \beta}(y) 
\end{eqnarray}
\end{subequations}
However, as we have shown in the present work, for a thermal state the temporal and spatial components of the projectors appear separately. Therefore, when written
in a covariant way,
the expressions quadratic in $h_{\mu\nu}$ become more complex combinations of contractions of the Riemann tensor with the four-velocity of the bath.
The replacements analog to those of Eq.\eqref{repP} read 
\begin{subequations}
\begin{eqnarray}
\label{proyectorescovariantes}
h_{00}(\bar{x})P_{00}P_{00}h_{00}(\bar{y})|\bar{p}|^4 &\rightarrow& 4
R_{00}(\bar{x}) R_{00}(\bar{y}) = 4 R_{\mu \nu}(\bar{x}) u^\mu u^\nu R_{\alpha \beta}(\bar{y}) u^\alpha u^\beta 
 \\
h_{00}(\bar{x})P_{00} P_{ij} h_{ij}(\bar{y}) |\bar{p}|^4 &\rightarrow& 
- 4 R_{00}(\bar{x}) R_{00}(\bar{y}) + 2 R_{00} (\bar{x}) R(\bar{y}) 
\nonumber \\
&=& 2 R_{\mu \nu}(\bar{x}) u^\mu u^\nu R(\bar{y}) - 4 R_{\mu \nu}(\bar{x}) u^\mu u^\nu R_{\alpha \beta}(\bar{y}) u^\alpha u^\beta 
\nonumber \\   
h_{ij}(\bar{x})P_{ij} P_{lm} h_{lm}(\bar{y}) |\bar{p}|^4 &\rightarrow& 
 \Big(- 2 R_{00}(\bar{x}) + R(\bar{x}) \Big) \Big(-2 R_{00}(\bar{y}) + R(\bar{y}) \Big)
\nonumber \\
&=& R(\bar{x}) R(\bar{y}) - 2 R(\bar{x}) R_{\mu \nu}(\bar{y}) u^\mu u^\nu - 2 R_{\mu \nu}(\bar{x}) u^\mu u^\nu R(\bar{y}) 
\nonumber \\
&& + 4 R_{\mu \nu}(\bar{x}) u^\mu u^\nu R_{\alpha \beta}(\bar{y}) u^\alpha u^\beta  \\
h_{ij}(\bar{x})P^{il} P^{jm} h_{lm}(\bar{y}) |\bar{p}|^4 &\rightarrow& 
R_{\mu \nu \alpha \beta}(\bar{x}) R^{\mu \nu \alpha \beta}(\bar{y}) - 4 R_{\mu \nu}(\bar{x}) u^\mu u^\nu R_{\alpha \beta}(\bar{y}) u^\alpha u^\beta - 8 R_{0i}(\bar{x}) R^{0i}(\bar{y})
\nonumber \\
&=& R_{\mu \nu \alpha \beta}(\bar{x}) R^{\mu \nu \alpha \beta}(\bar{y}) - 4 R_{\mu \nu}(\bar{x}) u^\mu u^\nu R_{\alpha \beta}(\bar{y}) u^\alpha u^\beta 
\nonumber \\
&& - 8 \Big( R_{\mu \nu}(\bar{x}) u^\mu u^\nu R_{\alpha \beta}(\bar{y}) u^\alpha u^\beta  + R_{\mu \nu}(\bar{x}) R_{\beta}^\nu(\bar{y}) u^\mu u^\beta \Big)
\nonumber \\
&=& R_{\mu \nu \alpha \beta}(\bar{x}) R^{\mu \nu \alpha \beta}(\bar{y}) - 12 R_{\mu \nu}(\bar{x}) u^\mu u^\nu R_{\alpha \beta} u^\alpha u^\beta \nonumber\\
&-& 8 R_{\mu \nu}(\bar{x}) R_{\beta}^\nu(\bar{y}) u^\mu u^\beta 
 \\
h_{0i}(\bar{x}) P^{00} P^{ij} h_{0j}(\bar{y}) |\bar{p}|^4 &\rightarrow& 4  R_{\mu \nu}(\bar{x}) u^\mu u^\nu R_{\alpha \beta}(\bar{y}) u^\alpha u^\beta + 4 R_{\mu \nu}(\bar{x}) R_{\beta}^\nu(\bar{y}) u^\mu u^\beta 
\end{eqnarray}
\end{subequations}

\section{Details of the  high temperature expansion}
\label{hightemperatureexpansion}

In this Appendix we present the high temperature expansion of the integrals  $I_{\alpha\gamma}$ and $\hat I_{\alpha\gamma}$ defined in Eqs.\eqref{Ialphagamma} and 
\eqref{hatI}, respectively.

\subsection{Real part of $I_{\alpha\gamma}$ }

The integral of interest is 
\begin{equation}
 I_{\alpha \gamma} = \int_0^\infty dz \frac{z^{\alpha+1}}{e^z - 1} I_\gamma, 
\end{equation}
where
\begin{eqnarray}\label{Igammadef} 
I_\gamma &=& \int_{-1}^1 d\chi\ Re \Bigg[  \frac{\chi^\gamma}{(z_0 + z\chi - i \epsilon)} \Bigg] \, \\
&=& \frac{1}{z} \left\{ \sum^{\gamma-1}_{k=0} \Bigg(- \frac{z_0}{z} \Bigg)^k \frac{[1 - (-1)^{\gamma - k}]}{(\gamma - k)} + \Bigg(- \frac{z_0}{z} \Bigg)^\gamma \Bigg[ \frac{1}{2} \log \Bigg(\frac{(z + z_0)^2}{(z - z_0)^2} \Bigg) \Bigg] \right\} \equiv \frac{1}{z} f_\gamma\left(\frac{z_0}{z}\right). \nonumber 
\end{eqnarray}
The high temperature expansion corresponds to the limit $z_0\to 0$.

The strategy to proceed with the evaluation is as follows, we will first split the integration range in two parts, and then in each part we will perform an series expansion of a certain factor of the integrand such that it is convergent for the values of $z$ in that range. Afterwards, the summation will be commuted with the integration, and, after further manipulations, a final expression in terms a power series in $z_0$ will be found. The idea is that many of the initial series expansions will be able to be resummed back to exact expressions of known functions.

The integrand is composed of two distinct factors, $z^\alpha/(e^z-1)$ and a function $f$ of the quotient $z_0/z$, so, since we are interested in finding an expansion for $z_0 \ll 1$, we shall split the integration range in $\int_0^\infty \rightarrow \int_0^1+\int_1^\infty$,
\begin{eqnarray}
I_{\alpha \gamma} = \int_0^\infty dz \frac{z^{\alpha+1}}{e^z - 1} I_\gamma 
= \int_0^1 dz \frac{z^\alpha}{e^z - 1} I_\gamma + \int_1^\infty dz \frac{z^\alpha}{e^z - 1} I_\gamma 
= I_{\alpha \gamma}^{(A)} + I_{\alpha \gamma}^{(B)}.
\end{eqnarray}
In the first segment we expand the first factor for $z \ll 1$,
\begin{equation}
 \frac{z^\alpha}{e^z-1} = z^{\alpha-1} \sum^{\infty}_{k=0} \frac{B_k}{k!} z^k, \label{Bernoulli-nums}
\end{equation}
where the $B_k$ coefficients are the well known Bernoulli numbers. In the second segment, the relation $z_0 < z$ always holds, and therefore we can expand the second factor (i.e the function f) in powers of $z_0/z$.

Let us now focus on the first part $I_{\alpha \gamma}^{(A)}$. Using Eq.~\eqref{Igammadef} and the expansion given in Eq.~\eqref{Bernoulli-nums}, we have
\begin{eqnarray}
I_{\alpha \gamma}^{(A)} &=& \int_0^1 dz \frac{z^\alpha}{e^z - 1} I_\gamma 
= \sum^{\infty}_{k=0} \frac{B_k}{k!} \int_0^1 dz z^{\alpha + k - 1} I_\gamma
\nonumber \\
&=& \sum^{\infty}_{k=0} \frac{B_k}{k!} \Bigg\{ \sum^{\gamma - 1}_{l=0} (-z_0)^l \frac{[1 - (-1)^{\gamma-l}]}{(\gamma - l)} \int_0^1 dz z^{\alpha+ k - l - 1} + (- z_0)^\gamma \Bigg[ \frac{1}{2} \int_0^1 dz z^{\alpha + k - \gamma - 1} \log \Bigg(\frac{(z + z_0)^2}{(z - z_0)^2} \Bigg) \Bigg]
 \Bigg\}
\nonumber \\
&=& \sum^{\gamma - 1}_{l=0} (-z_0)^l \frac{[1 - (-1)^{\gamma-l}] }{(\gamma - l)} \int_0^1 \frac{z^{\alpha - l}}{e^z - 1}
+ (-z_0)^\gamma \sum^\infty_{k=0} \frac{B_k}{k!} I_{L}^{(\tilde{\alpha} + k)} 
\nonumber \\
&\equiv& I_{\alpha \gamma}^{(A1)} + I_{\alpha \gamma}^{(A2)}, \label{IA}
\end{eqnarray}
where $\tilde{\alpha} = \alpha - \gamma$ and $I_{\alpha \gamma}^{(A1)}$ and $I_{\alpha \gamma}^{(A2)}$ refer to the first and second term respectively. Also, $I_L^{\lambda}$ can be written as a function of $z_0$ for a generic integer $\lambda > 0$ as follows
\begin{eqnarray}
I_L^{(\lambda)} &=& \frac{1}{2} \int_0^1 dz\ z^{\lambda -1} \log \Bigg[\frac{(z + z_0)^2}{(z - z_0)^2} \Bigg] 
\nonumber \\
&=& \frac{1}{\lambda} \Bigg\{ \sum^{\lambda - 1}_{l=0} [1 - (-1)^l] \frac{z_0^l}{(\lambda - l)} - [1 - (-1)^\lambda] z_0^\lambda \log(z_0) 
\nonumber \\
&& \ \ \ \ + [1 - (-z_0)^l] \log(1 + z_0) - (1 - z_0^\lambda) \log(1 - z_0) \Bigg\},
\label{Ilambda}
\end{eqnarray}
however, notice in Eq.~\eqref{IA} that we also need $I_L^{(0)} = \frac{\pi^2}{2} - \sum_{l=1}^{\infty} [1 - (-1)^l] z_0^l/l^2$. This will force us to treat the cases $\tilde{\alpha} > 0$ and $\tilde{\alpha} = 0$ separately.

For the first case, $\tilde{\alpha} > 0$, using Eq.~\eqref{Ilambda} we have
\begin{eqnarray}
I_{\alpha \gamma}^{(A2)} = (- z_0)^\gamma &\Biggl\{& \sum^\infty_{k=0} \frac{B_k}{k!} \frac{1}{(\tilde{\alpha} + k)} \sum^{\tilde{\alpha} - 1 + k}_{l=0} [1 - (-1)^l] \frac{z_0^l}{(\tilde{\alpha} + k - l)}
\nonumber \\
&& - \Bigg( \sum^\infty_{k=0} \frac{B_k}{k!} \frac{z^{\tilde{\alpha} + k}}{(\tilde{\alpha} + k)} \Bigg) \log\left( \frac{z_0}{1 - z_0} \right) + \Bigg( \sum^\infty_{k=0} \frac{B_k}{k!} \frac{(-z_0)^{\tilde{\alpha} + k}}{(\tilde{\alpha} + k)} \Bigg) \log\left(\frac{z_0}{1+z_0}\right)
\nonumber \\
&& + \Bigg( \sum^\infty_{k=0} \frac{B_k}{k!} \frac{1}{(\tilde{\alpha} + k)} \Bigg) \log \Bigg( \frac{1+z_0}{1- z_0} \Bigg) \Biggr\}.
\end{eqnarray}
The first line is the one that needs to be worked out the most. The idea is swap the summation order to be able to then resum the series of the Bernoulli coefficients. Schematically,
\begin{eqnarray}
\sum^\infty_{k=0} \sum^{\tilde{\alpha}-1 + k}_{l=0} \longrightarrow 
\sum^{\infty}_{k=0} \Bigg(\sum^{k-1}_{l=0} + \sum^{\tilde{\alpha}+k-1}_{l=k} \Bigg)
=
\sum^{\infty}_{l=0}\  \sum^{\infty}_{k=l+1} + \sum^\infty_{k=0}\  \sum^{\tilde{\alpha}-1}_{n=0}.
\end{eqnarray}
After some further manipulations, we arrive at 
%
%
%
\begin{eqnarray}
I_{\alpha \gamma}^{(A2)} = (-z_0)^\gamma &\Biggl\{& \sum^\infty_{l=0} [1 - (-1)^l] \frac{z_0^l}{l} \Bigg( \int^1_0 dz \frac{z^{\tilde{\alpha}-l+\Delta}}{e^z - 1} - \sum^\infty_{k=0} \frac{B_k}{k!} \frac{1}{(\tilde{\alpha} + k)} \Bigg)
\nonumber \\
 && - \sum^\infty_{l=0} [1 - (-1)^l] z_0^l \sum_{k=0}^l \frac{B_k}{k!} \frac{1}{(\tilde{\alpha} + k - l + \Delta) (\tilde{\alpha} + k + \Delta)}
\nonumber \\
&& + \Bigg(\sum^\infty_{k=0} \frac{B_k}{k!} \frac{z_0^k}{(\tilde{\alpha} + k )} \Bigg) \Bigg( \sum^{\tilde{\alpha} - 1}_{n=0} \frac{z_0^n}{(\tilde{\alpha} - n)} \Bigg)
- \Bigg(\sum^\infty_{k=0} \frac{B_k}{k!} \frac{(-z_0)^k}{(\tilde{\alpha} + k)} \Bigg) \Bigg( \sum^{\tilde{\alpha} - 1}_{n=0} \frac{(-z_0)^n}{(\tilde{\alpha} - n)} \Bigg) 
\nonumber \\
&& - \Bigg( \sum^\infty_{k=0} \frac{B_k}{k!} \frac{z^{\tilde{\alpha} + k}}{(\tilde{\alpha} + k)} \Bigg) \log\left( \frac{z_0}{1 - z_0} \right) + \Bigg( \sum^\infty_{k=0} \frac{B_k}{k!} \frac{(-z_0)^{\tilde{\alpha} + k}}{(\tilde{\alpha} + k)} \Bigg) \log\left(\frac{z_0}{1 + z_0} \right) 
\nonumber \\
&& + \Bigg( \sum^\infty_{k=0} \frac{B_k}{k!} \frac{1}{(\tilde{\alpha} + k)} \Bigg) \log \Bigg( \frac{1+z_0}{1- z_0} \Bigg) \Biggr\},
\end{eqnarray}
where $\Delta$ has been introduced as a regulator in order to avoid spurious divergences that have been introduced by the splitting. These divergences will not be present when recombining everything in the final result, for which the limit $\Delta \to 0$ can be safely taken.

For the case $\tilde{\alpha} = 0$, we have $\alpha = \gamma$ and thus
\begin{eqnarray}
I_{\alpha \alpha}^{(A2)} &=& (-z_0)^\gamma \sum^\infty_{k=0} \frac{B_k}{k!} I_L^{(k)}
= (-z_0)^\gamma \Bigg[ I_L^{(0)} + \sum^\infty_{k=1} \frac{B_k}{k!}  I_L^{(k)} \Bigg]
\nonumber \\
&=& (- z_0)^\gamma \Bigg[ \frac{\pi^2}{2} - \sum^\infty_{l=1} [1 - (-1)^l] \frac{z_0^l}{l^2} 
+ \sum^\infty_{k=1} \frac{B_k}{k!}  I_L^{(k)}\Bigg], 
\end{eqnarray}
where now
\begin{eqnarray}
\sum^{\infty}_{k=1} \frac{B_k}{k!} I_L^{(k)} &=& \sum^\infty_{k=1} \frac{B_k}{k!} \frac{1}{k} \sum^{k-1}_{l=1} [1 - (-1)^l] \frac{z_0^l}{(k-l)} 
- \Bigg( \sum^\infty_{k=1} \frac{B_k}{k!} \frac{z^k}{k}\Bigg) \log \Bigg(\frac{z_0}{1 - z_0}\Bigg)
\nonumber \\
&+& \Bigg( \sum^\infty_{k=1} \frac{B_k}{k!} \frac{(-z_0)^k}{k}\Bigg) \log \Bigg(\frac{z_0}{1+z_0} \Bigg) + \Bigg( \sum^\infty_{k=1} \frac{B_k}{k!} \frac{1}{k}\Bigg) \sum^{\infty}_{l=1} [1 - (-1)^l] \frac{z_0^l}{l}. 
\end{eqnarray}
We then have
\begin{eqnarray}
I_{\alpha \alpha}^{(A2)} &=& \Bigg\{ \frac{\pi}{2} + \sum^\infty_{l=1} [1 - (-1)^l] \frac{z_0^l}{l} \int^1_0 dz \frac{z^{\Delta - l}}{e^z - 1} 
-  \sum^\infty_{l=1} [1 - (-1)^l] z_0^l \sum^l_{k=1} \frac{B_k}{k!} \frac{1}{(k - l + \Delta) (k + \Delta)} 
\nonumber \\
&& - \Bigg( \sum^\infty_{k=1} \frac{B_k}{k!} \frac{z_0^k}{k} \Bigg) \log \Bigg(\frac{z_0}{1 - z_0} \Bigg) + \Bigg( \sum^\infty_{k=1} \frac{B_k}{k!} \frac{(-z_0)^k}{k}\Bigg)  \log \Bigg( \frac{z_0}{1 + z_0} \Bigg) \Bigg\} (-z_0)^\gamma
\end{eqnarray}

The final part we need is $I_{\alpha \gamma}^{(B)}$. Similarly as before,
\begin{eqnarray}
I_{\alpha \gamma}^{(B)} &=& \int^\infty_1 dz \frac{z^{\alpha+1}}{e^z - 1} I_\gamma
\nonumber \\
&=& \int^\infty_1 dz \frac{z^{\alpha}}{e^z - 1} \Bigg\{ \sum^{\gamma - 1}_{k=0} \Bigg(- \frac{z_0}{z} \Bigg)^k \frac{[1 - (-1)^{\gamma - k}]}{(\gamma - k)} + \Bigg(- \frac{z_0}{z} \Bigg)^\gamma \Bigg[ \sum^{\infty}_{l=0} \frac{[1 - (-1)^l]}{l} \Bigg(\frac{z_0}{z} \Bigg)^l \Bigg] \Bigg\}
\nonumber \\
&\equiv& I_{\alpha \gamma}^{(B1)} + I_{\alpha \gamma}^{(B2)},
\end{eqnarray}

Finally, we put everything in the following way
\begin{eqnarray}
I_{\alpha \gamma}^{(A1)} + I_{\alpha \gamma}^{(B1)} = \sum^{\gamma - 1}_{l=1} (-z_0)^l \frac{[1 - (-1)^{\gamma - l }]}{(\gamma - l)} \Gamma[\alpha - l] \zeta[\alpha - l], \label{part1}
\end{eqnarray}
where $\zeta(s)$ and $\Gamma(s)$ are the Riemann Zeta and Euler Gamma functions respectively. For the other contributions we consider the two cases one by one. For $\tilde{\alpha}>0$ we have
\begin{eqnarray}
I_{\alpha \gamma}^{(A2)} + I_{\alpha \gamma}^{(B2)} = (-z_0)^\gamma &\Biggl\{& \sum^\infty_{l=1} [1 - (-1)^l] \frac{z_0^l}{l} \Gamma[\tilde{\alpha} - l + 1 + \Delta] \zeta[\tilde{\alpha} - l + 1 + \Delta] 
\nonumber \\
&& - \sum^\infty_{l=1} [1 - (-1)^l] z_0^l \sum^l_{k=0} \frac{B_k}{k!} \frac{1}{(\tilde{\alpha} + k - l + \Delta) (\tilde{\alpha} + k + \Delta)}
\nonumber \\
&& + \Bigg( \sum^\infty_{k=0} \frac{B_k}{k!} \frac{z_0^k}{(\tilde{\alpha} + k )} \Bigg) \Bigg[ \sum^{\tilde{\alpha} - 1}_{n=0} \frac{z_0^n}{(\tilde{\alpha} - n)} - z_0^{\tilde{\alpha}} \log \Bigg(\frac{z_0}{1 - z_0} \Bigg) \Bigg] \label{part2not0}\\
&& - \Bigg( \sum^\infty_{k=0} \frac{B_k}{k!} \frac{(-z_0)^k}{(\tilde{\alpha} + k )} \Bigg) \Bigg[ \sum^{\tilde{\alpha} - 1}_{n=0} \frac{(-z_0)^n}{(\tilde{\alpha} - n)} - (-z_0)^{\tilde{\alpha}} \log \Bigg(\frac{z_0}{1 + z_0} \Bigg) \Bigg] \Biggr\}, \nonumber
\end{eqnarray}
while for $\tilde{\alpha} = 0$, we obtain
\begin{eqnarray}
I_{\alpha \alpha}^{(A2)} + I_{\alpha \alpha}^{(B2)} = (- z_0)^\gamma &\Bigg\{& \frac{\pi^2}{2} + \sum^\infty_{l=1} [1 - (-1)^l] \frac{z_0^l}{l} \Gamma[1 - l + \Delta] \zeta[1 - l + \Delta] 
\nonumber \\
&&-  \sum^{\infty}_{l=1} [1 - (-1)^l] z_0^l \sum^l_{k=1} \frac{B_k}{k!} \frac{1}{(k - l + \Delta) (k + \Delta)} \label{part20} \\
&&- \Bigg( \sum^\infty_{k=1} \frac{B_k}{k!} \frac{z_0^k}{k} \Bigg) \log \Bigg(\frac{z_0}{1 - z_0} \Bigg) + \Bigg(\sum^\infty_{k=1} \frac{B_k}{k!} \frac{(-z_0)^k}{k}\Bigg) \log \Bigg( \frac{z_0}{1+z_0} \Bigg) \Bigg\} \nonumber
\end{eqnarray}

The final result is obtained by adding Eq.~\eqref{part1} with Eq.~\eqref{part2not0} when $\alpha > \gamma$ or with Eq.~\eqref{part20} when $\alpha = \gamma$. To achieve a given order in powers of $z_0$, the formal expressions must be truncated at a given order which depends on the particular values of $\alpha$ and $\gamma$. At the end, the limit $\Delta \to 0$ is to be taken.

Recalling that $z_0 = \beta |\bar{p}|/2$, we now show the results for the particular values of $\alpha$ and $\gamma$ needed, up to $\mathcal{O}(\beta^0)$,
\begin{eqnarray}
\label{expansionies}
I_{20} &=& \frac{\pi ^2}{12 \beta ^2 |\bar{p}|^2}+\frac{1}{288} \left[6 \log \left(\frac{\beta  |\bar{p}|}{2} \right) + 6 \gamma -2 - 6 \log (2 \pi )\right]
\nonumber \\
I_{22} &=& -\frac{\pi ^2}{12 \beta ^2 |\bar{p}|^2}+\frac{\pi ^2}{16 \beta  |\bar{p}|} + \frac{1}{16} \left[\log \left(\frac{\beta  |\bar{p}|}{2}\right) + \gamma - 1 - \log (2 \pi )\right]
\nonumber \\
I_{31} &=& \frac{\pi ^4}{15 \beta ^4 |\bar{p}|^4}-\frac{\pi ^2}{24 \beta ^2 |\bar{p}|^2} + \frac{1}{576} \left[-6 \log \left(\frac{\beta  |\bar{p}|}{2}\right)-6 \gamma+ 2 + 6 \log (2 \pi )\right]
\nonumber \\
I_{33} &=& \frac{\pi ^4}{45 \beta ^4 |\bar{p}|^4}+\frac{\pi ^2}{24 \beta ^2 |\bar{p}|^2}-\frac{\pi ^2}{32 \beta  |\bar{p}|}+\frac{1}{32} \left[\log \left(\frac{4 \pi}{\beta  |\bar{p}|}\right) - \gamma + 1\right]
\nonumber \\
I_{40} &=& \frac{\pi^4}{30 \beta^4 |\bar{p}|^4} + \frac{\pi^2}{144 \beta^2 |\bar{p}|^2} + \frac{1}{14400} \Bigg[45 \log \left(\frac{\beta  |\bar{p}|}{2}\right)+45 \gamma - 9 - 45 \log (2 \pi )\Bigg]
\nonumber \\
I_{42} &=& -\frac{\pi ^4}{30 \beta ^4 |\bar{p}|^4}+\frac{\pi ^2}{48 \beta ^2 |\bar{p}|^2}+\frac{1}{1152} \Bigg[6 \log \left(\frac{\beta  |\bar{p}|}{2}\right) + 6 \gamma - 2 - 6 \log (2 \pi ) \Bigg]
\nonumber \\
I_{44} &=&  -\frac{\pi ^4}{90 \beta ^4 |\bar{p}|^4} - \frac{\pi ^2}{48 \beta ^2 |\bar{p}|^2} + \frac{\pi ^2}{64 \beta  |\bar{p}|} + \frac{1}{64} \left[ \log \left(\frac{\beta  |\bar{p}|}{2}\right) + \gamma -1 - \log (2 \pi )\right]
\end{eqnarray}

\subsection{Imaginary part of $\hat I_{\alpha\beta}$}

The  imaginary part of  $\hat{I}_{\alpha \beta}$ reads
\begin{eqnarray}
(2 \pi) Im\Big[\hat{I}_{\alpha \beta}(|\bar{p}|)\Big] &=& (2\pi) \int_0^\infty dk\ k^{\alpha+1} \frac{1}{\vert \bar p\vert^\alpha} \Big(n(k) + n(k)^2 \Big) \int_{-1}^1 dx\ x^\gamma\ Im\Bigg[\frac{1}{k^2+2 k |\bar{p}| x - i \epsilon} \Bigg]
\nonumber \\
&=& \frac{\pi^2 (-1)^\gamma}{2^\gamma (\vert\bar p\vert\beta)^{\alpha + 1 - \gamma}} \int^\infty_{z_0} dz\ z^{\alpha+\gamma } \Bigg[ \frac{1}{e^z-1} + \frac{1}{(e^z - 1)^2} \Bigg] \label{parteim}
\end{eqnarray}•

In order to expand this expression in powers of $z_0 (\beta\to 0)$, we define
\begin{equation}
f(z) = z^{\alpha+\gamma } \Bigg[ \frac{1}{e^z-1} + \frac{1}{(e^z - 1)^2} \Bigg]\, ,
\end{equation}•
and split the integral in Eq. (D18)  as
\begin{eqnarray}
 \int^\infty_{z_0} dz\ f(z)
&=& \int_0^\infty dz\ f(z) - \sum^\infty_{k=0} \frac{\partial^k f}{\partial z^k}\Bigg|_{z_0=0}  \frac{z_0^{k+1}}{(k+1)!}\, .
\end{eqnarray}•

Then
\begin{eqnarray}
&& Im\Bigg( \frac{\hat{I}_{22} }{4} + \hat{I}_{31} + \hat{I}_{40} \Bigg) =  \frac{2 \pi ^5}{15 \beta ^5 |\bar{p}|^5} - \frac{\pi ^3}{12 \beta ^3 |\bar{p}|^3} + \frac{11 \pi }{96 \beta ^2 |\bar{p}|^2} - \frac{\pi }{64 \beta |\bar{p}|} + \frac{9\pi }{10240}
\nonumber \\
&& Im\Bigg( - \frac{\hat{I}_{22}}{2} - \frac{\hat{I}_{31}}{2} - \frac{\hat{I}_{33}}{2} + \hat{I}_{40} - \hat{I}_{42} \Bigg) = \frac{2 \pi ^5}{15 \beta ^5 |\bar{p}|^5} - \frac{\pi }{96 \beta ^2 |\bar{p}|^2} + \frac{\pi }{64 \beta  |\bar{p}|} - \frac{119 \pi }{92160}
\nonumber \\
&& Im\Bigg( \frac{\hat{I}_{22}}{4} - \frac{\hat{I}_{31}}{2} + \frac{\hat{I}_{33}}{2} + \frac{\hat{I}_{40}}{8} - \frac{\hat{I}_{42}}{4} + \frac{\hat{I}_{44}}{8} \Bigg)  = \frac{\pi ^5}{60 \beta ^5 |\bar{p}|^5} + \frac{\pi ^3}{32 \beta ^3 |\bar{p}|^3} - \frac{37 \pi }{768 \beta ^2 |\bar{p}|^2} - \frac{\pi }{512 \beta |\bar{p}|} + \frac{241 \pi }{737280}
\nonumber\\
&& Im\Bigg( \frac{\hat{I}_{40}}{4} - \frac{\hat{I}_{42}}{2} + \frac{\hat{I}_{44}}{4} \Bigg) =  \frac{\pi ^5}{30 \beta ^5 |\bar{p}|^5} - \frac{\pi ^3}{48 \beta ^3 |\bar{p}|^3} + \frac{11 \pi }{384 \beta ^2 |\bar{p}|^2} - \frac{\pi }{256 \beta |\bar{p}|} + \frac{9 \pi }{40960}
\nonumber \\
&& Im\Bigg(\hat{I}_{40} - \hat{I}_{42} \Bigg) =  \frac{2 \pi ^5}{15 \beta ^5 |\bar{p}|^5} - \frac{\pi ^3}{24 \beta ^3 |\bar{p}|^3} + \frac{5 \pi }{96 \beta ^2 |\bar{p}|^2} - \frac{19 \pi }{92160}
\end{eqnarray}•

\section{Details of the conformal transformation of the ultrastatic effective action}\label{app-conformal}

Consider the conformal transformation of Eq.~\eqref{conformal-transf} in four dimensions. The relevant transformation rules for transforming $\Gamma[\bar{g}_{\mu\nu}]$ are (see Ref.~\cite{Birrel}), p.38)
\begin{subequations} \label{conformal-transf-rules}
\begin{eqnarray} 
 R^{\nu}_{\,\,\mu} &=& \Omega^{-2} \left[ \bar{R}^{\nu}_{\,\,\mu} + 2 \Omega \bar{\nabla}^{\nu} \bar{\nabla}_{\mu} (\Omega^{-1}) - \frac{1}{2} \delta^{\nu}_{\,\,\mu} \Omega^{-2} \bar{\square}(\Omega^2) \right], \\
 R &=& \Omega^{-2} \left( \bar{R} + 2 \Omega \bar{\square} (\Omega^{-1}) - 2 \Omega^{-2} \bar{\square}(\Omega^2) \right) \equiv \Omega^{-2} \left( \bar{R} - \Delta \bar{R} \right), \label{conformal-R-transf}\\ 
 \phi &=& \Omega^{-1} \bar{\phi}, \label{conformal-phi-transf}
\end{eqnarray}
\end{subequations}
where all the quantities with an overbar are evaluated on the ultrastatic metric $\bar{g}_{\mu\nu}$. It is also useful to remember that a conformally coupled and massless field is conformally invariant, hence
\begin{equation}
 \left(-\square + \frac{R}{6} \right) \phi = \Omega^{-3/2} \left(-\bar{\square} + \frac{\bar{R}}{6} \right) \bar{\phi},
\end{equation}
and thus, the propagator for a conformally invariant field transforms as
\begin{equation}
 G(x,x') = \Omega^{-1}(x) \bar{G}(x,x') \Omega^{-1}(x),
\end{equation}
which can also be inferred from the transformation of $\phi$, Eq.~\eqref{conformal-phi-transf}.

With these transformation rules, it is straightforward to check that a more generic, non conformally-invariant field with mass $m$ and coupling to the curvature $\xi$, will see its inverse propagator change in the following way,
\begin{equation}
 \left(-\square + m^2 + \xi R \right) \phi = \Omega^{-3/2} \left[-\bar{\square} + \Omega^2 m^2 + \xi \bar{R} - \left( \xi - \frac{1}{6} \right) \Delta \bar{R} \right] \bar{\phi}.
\end{equation}

In order to profit from the known results for ultrastatic metrics of Ref.~\cite{GusevZelnikov}, where they find the thermal effective action for operators of the form $-\bar{\square} - \bar{P} + \frac{\bar{R}}{6}$, we choose the potential $\bar{P}$ to be
\begin{equation}
 \bar{P} = - \Omega^2 m^2 - \left( \xi - \frac{1}{6} \right) \left( \bar{R} - \Delta\bar{R} \right) = - \Omega^2 \left[ m^2 + \left( \xi - \frac{1}{6} \right) R \right],
\end{equation}
where in the second equality we used the rule \eqref{conformal-R-transf} to find an expression in terms of quantities defined for the static metric $g_{\mu\nu}$. The same tactic is to be applied to the whole ultrastatic effective action $\Gamma[\bar{g_{\mu\nu}},\Omega]$ in order to find its static counterpart $\Gamma[g_{\mu\nu}]$. 

Since we are working with the weak field approximation, it is useful to consider the linearized versions of some of the rules \eqref{conformal-transf-rules},
\begin{subequations} 
\begin{eqnarray} 
 R^{\nu}_{\,\,\mu} &=& \bar{R}^{\nu}_{\,\,\mu} + \partial^{\nu} \partial_{\mu} h_{00} + \frac{\delta^{\nu}_{\,\,\mu}}{2} \nabla^2 h_{00}, \\
 R &=& \bar{R} + 3 \nabla^2 h_{00}.
\end{eqnarray}
\end{subequations}
For the ultrastatic metric we have $\bar{h}_{00}=0$. However, it is through the 
$\Omega$'s coming from both $\bar{P}$ and the trasformation rules, that $h_{00}$ contributions arise. These must be such that when combined with the rest of the elements they form  geometrical objects associated with the static metric $g_{\mu\nu}$.

\section*{Acknowledgments}
This research was supported by Agencia Nacional de Promoci\'on Cient\'ifica y Tecnol\'ogica (ANPCyT), Consejo Nacional de Investigaciones
Cient\'ificas y T\'ecnicas (CONICET), and Universidad Nacional de Cuyo (UNCuyo).

\end{document}